\documentclass[twocolumn, twocolappendix]{aastex701}

\usepackage{caption}
\usepackage{amsmath}

\begin{document}

\title{Interplanetary Dust  Ablation Seeds Silicate Clouds in Exoplanet Atmospheres}

\author[orcid=0000-0003-1285-3433,sname='Kiefer']{Sven Kiefer}
\affiliation{Department of Astronomy, University of Texas at Austin, 2515 Speedway, Austin, TX 78712, USA}
\affiliation{Center for Planetary Systems Habitability, The University of Texas at Austin, Austin, TX, USA}
\email[show]{sven.kiefer@utexas.edu}  

\author[orcid=0000-0002-4043-0377,sname='Gosselin']{Gregory J. Gosselin} 
\affiliation{Center for Planetary Systems Habitability, The University of Texas at Austin, Austin, TX, USA}
\affiliation{Institute for Geophysics, The University of Texas at Austin, 1100 Burnet road BLDG ROC, Austin, TX 78758, USA}
\email{gregory.gosselin@jsg.utexas.edu}

\author[orcid=0000-0002-4404-0456,sname='Morley']{Caroline V. Morley} 
\affiliation{Department of Astronomy, University of Texas at Austin, 2515 Speedway, Austin, TX 78712, USA}
\affiliation{Center for Planetary Systems Habitability, The University of Texas at Austin, Austin, TX, USA}
\email{cmorley@utexas.edu}  

\author[orcid=0000-0003-4740-9068,sname='Gulick']{Sean P. S. Gulick } 
\affiliation{Center for Planetary Systems Habitability, The University of Texas at Austin, Austin, TX, USA}
\affiliation{Institute for Geophysics, The University of Texas at Austin, 1100 Burnet road BLDG ROC, Austin, TX 78758, USA}
\affiliation{Department of Earth and Planetary Sciences, Jackson School of Geosciences, The University of Texas at Austin, Austin, TX, USA}
\email{sean@ig.utexas.edu}  




\begin{abstract}
  Interplanetary dust particles (IDPs) are present throughout the Solar System, with Earth experiencing a mass flux of roughly 100 tonnes of IDPs per day. These particles ablate in the upper atmosphere, delivering aerosols that form noctilucent and stratospheric clouds. Since IDPs are present in exoplanet systems as exozodiacal dust disks (exozodis), it is likely that exoplanets experience a constant infall of IDPs as well.
  In this work, we investigate the effect of IDPs on the cloud structure of gas-giant exoplanets and brown dwarfs, and determine their observability.
  We apply the microphysical cloud model \texttt{Nimbus} to the Sonora Diamondback model grid to cover a wide range of effective temperatures, surface gravities, atmospheric mixing constants, and IDP mass fluxes.
  Our results show that IDP infall increases cloud particle number densities, decreases cloud particle radii, and enhances silicate features in transmission and thermal emission spectra. Exoplanets with lower atmospheric mixing rates, effective temperatures, and gravity are more sensitive to IDPs. Most notably, they lead to observable silicate absorption features in colder planets where silicate clouds would otherwise not be observable. WASP-107~b is an example of such an exoplanet with observational evidence of silicate particles. Using \texttt{Nimbus}, we found that a 100 times Earth-like IDP mass flux with a Solar-System-like IDP composition of quartz (SiO$_2$), enstatite (MgSiO$_3$), and iron (Fe) can explain the observations of high altitude silicate particles in WASP-107~b. 
  
\end{abstract}



\section{Introduction} 
\label{sec:Introduction}

Interplanetary Dust Particles (IDPs) are micron- to centimetre-sized particles that originate from asteroids and comets \citep{koschny_interplanetary_2019}. These particles naturally disperse, leading to a Solar System wide distribution \citep{plane_impacts_2017, poppe_improved_2016}. Earth receives roughly 100 tonnes of IDPs per day \citep{cplane_cosmic_2012} which ablate around 100~km altitude \citep{cziczo_ablation_2001} and seed the formation of stratospheric or noctilucent clouds by delivering refractive species to the upper atmosphere \citep{biermann_unsuitability_1996, voigt_nitric_2005, bardeen_numerical_2008, cplane_cosmic_2012, bardeen_improved_2013, hartwick_high-altitude_2019, tritscher_polar_2021, james_importance_2023}. IDP infall is also thought to be a source of organic material on Mars \citep{moores_uv_2012, frantseva_delivery_2018}. Planet formation is thought to result in remnant planetesimal populations like the asteroid and Kuiper belts in our Solar System. Those undergo collisional cascades for billions of years, continuously generating interplanetary dust \citep{hughes_debris_2018}. IDPs have also been shown to be present in extra solar systems \citep{mennesson_constraining_2014, defrere_first-light_2015, defrere_hosts_2021, garreau_hosts_2025}. Multiple studies have investigated the origins \citep{marboeuf_extrasolar_2016, faramaz_inner_2017, rigley_comet_2022} and distributions \citep{kennedy_warm_2015} of extrasolar IDPs, especially in the form of exozodiacal dust disks (exozodis) \citep{wyatt_theory_2025}. Understanding these particles is important since they can obscure habitable planets in direct imaging \citep{defrere_nulling_2010, roberge_exozodiacal_2012, stark_maximizing_2014, kennedy_warm_2015, currie_exozodi_2026} or they can fall onto planets, affecting their atmospheric structure.

Gas-giant exoplanets and brown dwarfs have extended hydrogen-helium atmospheres, which are favourable for observations with space-based telescopes like Hubble (HST), Spitzer, and James Webb (JWST). Observations with these telescopes have led to the detection of various molecules \citep[e.g.][]{tsai_photochemically_2023, feinstein_early_2023, rustamkulov_early_2023, bell_nightside_2024}, atmospheric inhomogeneities \citep[e.g.][]{espinoza_inhomogeneous_2024, murphy_evidence_2024, steinrueck_limb_2025, chen_asymmetry_2025}, and silicate clouds \citep[e.g.][]{grant_jwst-tst_2023, dyrek_so2_2023, molliere_evidence_2025, hoch_silicate_2025, suarez_ultracool_2022, suarez_ultracool_2023, miles_jwst_2023, welbanks_high_2024}. Here we focus on the detection of silicate clouds within exoplanet atmospheres as their primary materials (enstatite, forsterite, and quartz) are found in Solar System chondrites and IDPs \citep{thomas_carbon_1993, keller_carbon_1993, lodders_relative_2021}. While it is also important to investigate the effect of IDPs on terrestrial planets, current telescopes are not precise enough to gain observational insights into these types of exoplanets \citep[see e.g.][]{may_double_2023, lustig-yaeger_jwst_2023, moran_high_2023, scarsdale_jwst_2024, kirk_jwstnircam_2024, alderson_jwst_2024, glidden_jwst-tst_2025, xue_jwst_2025, kreidberg_first_2025}. For this work, we thus focus on hot gas-giant exoplanets and brown dwarfs.

\begin{figure}
    \centering
    \includegraphics[width=\linewidth]{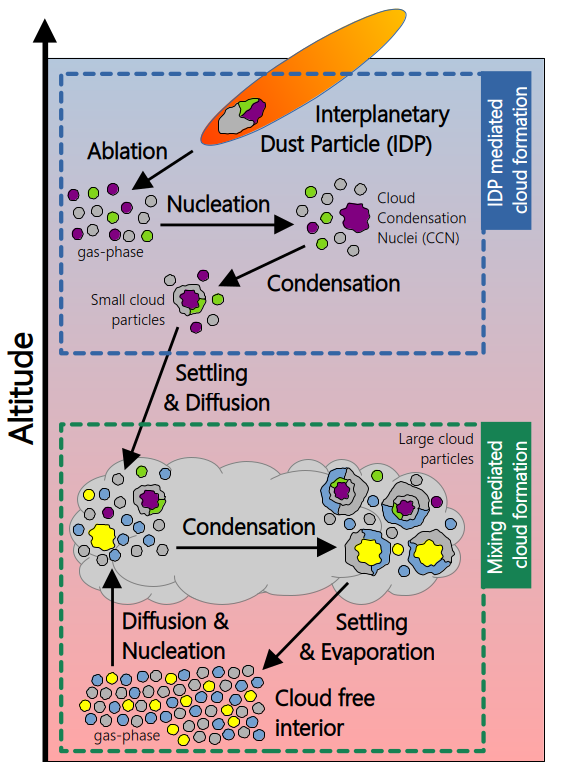}
    \caption{Connection between Interplanetary Dust Particles (IDPs) and cloud formation in planetary atmospheres. The ablation products of IDPs nucleate and condense at high altitudes, providing Cloud Condensation Nuclei (CCNs) for cloud materials diffusing from the planetary interior. The additional source of CCNs affects cloud particle radii, number densities, and Mass Mixing Ratios (MMRs). The differently coloured particles represent cloud particle materials.}
    \label{fig:illust}
\end{figure}

\begin{figure}
    \centering
    \includegraphics[width=\linewidth]{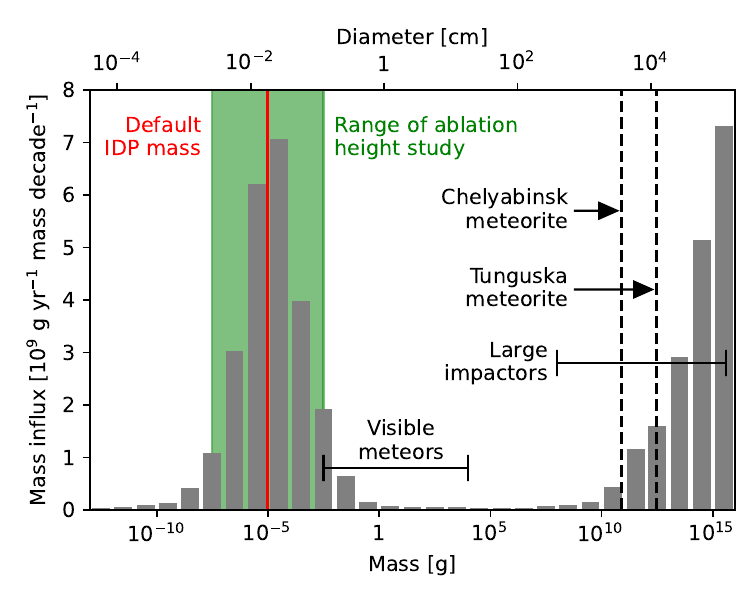}
    \caption{Mass influx per particle size. Data taken from  \citet{kyte_accretion_1986}, \citet{love_direct_1993}, \citet{flynn_extraterrestrial_2002}, and \citet{cplane_cosmic_2012}. The sizes of the Chelyabinsk \citep{popova_chelyabinsk_2013} and Tunguska \citep{kulik_results_1940} meteorites are shown for comparison. For reference, the Chicxulub meteor which caused the Cretaceous–Paleogene extinction event had a diameter of $\sim12~\mathrm{km}$ \citep[][]{gulick_end_2025}.}
    \label{fig:Idp_mass_dist}
\end{figure}

Cloud formation in gaseous atmospheres is governed by three mechanisms: nucleation, growth, and gravitational settling (Figure~\ref{fig:illust}). Nucleation is the first step of cloud formation, describing the clustering of molecules until they become large enough to act as cloud condensation nuclei (CCNs) \citep{patzer_dust_1998,  lee_dust_2015, bromley_under_2016,  kohn_dust_2021, sindel_revisiting_2022, gobrecht_bottom-up_2022, gobrecht_bottom-up_2023, kiefer_effect_2023, lecoq-molinos_vanadium_2024}. The rate at which materials nucleate determines the volume mixing ratio (VMR) of cloud particles. Once CCNs are present, other species can grow onto them, leading to larger particles and increasing the mass mixing ratio (MMR) of cloud particles \citep{ackerman_precipitating_2001, helling_dust_2006, helling_sparkling_2019, kiefer_fully_2024}. The major cloud forming materials in hot Jupiters and L-dwarfs are enstatite (MgSiO$_3$), forsterite (Mg$_2$SiO$_4$), quartz (SiO$_2$) and iron (Fe) \citep{gao_aerosol_2020}, the same materials as found in Solar System IDPs. Once cloud particles have grown larger, they gravitationally settle deeper into the atmosphere \citep{woitke_dust_2003}. IDPs occupy an interesting place in the cloud formation cycle by delivering cloud forming materials to the upper atmosphere (Figure~\ref{fig:illust}). It is therefore possible that IDPs have an observable effect on the atmospheric structure of exoplanets \citep{arras_dust_2022, madurowicz_infrared_2023}. 

In this work, we study the effect of IDPs on the cloud structures of gas-giant exoplanets and investigate their observability through transmission and thermal emission spectra. To achieve this, we estimate the ablation height and mass flux of IDPs (Section~\ref{sec:ablation}), advance the exoplanet cloud model \texttt{Nimbus} to account for IDP infall (Section~\ref{sec:nimbus}), and produce synthetic observations with the following software packages: \texttt{PICASO} for radiative transfer calculations \citep{batalha_exoplanet_2019, mukherjee_picaso_2023, mang_picaso_2026}, \texttt{Virga} for modelling the spectral impact of clouds \citep{batalha_condensation_2026, moran_fractal_2025}, and \texttt{MieNet} to rapidly compute mixed-cloud optical constants. We apply \texttt{Nimbus} to the Sonora Diamondback atmospheric model grid \citep{morley_sonora_2024} to determine which exoplanets are most affected by IDP infall. For irradiated planets observed in transmission, we use \texttt{Picaso} to calculate atmospheric structures covering the same parameter space. The hot Jupiter WASP-107~b \citep{anderson_discoveries_2017} is of particular interest for this study, since high altitude silicate particles have been observed in its atmosphere \citep{dyrek_so2_2023, welbanks_high_2024}. In this work, we investigate if these observations can be explained through IDP infall (Section~\ref{sec:obs}).

\section{Ablation of interplanetary dust particles in hydrogen dominated atmospheres}
\label{sec:ablation}

\begin{figure}
    \centering
    \includegraphics[width=\linewidth]{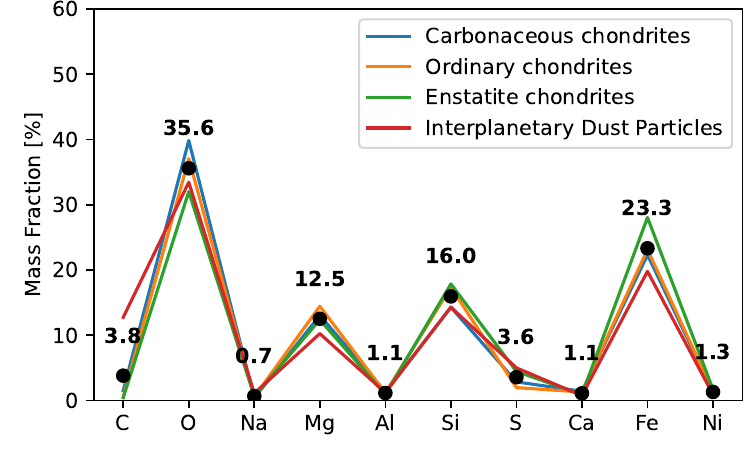}
    \caption{Average composition of different interplanetary materials. Carbonaceous, ordinary, and Enstatite chondrite data are from \citet{lodders_relative_2021} and IDP data are from \citet{thomas_carbon_1993} and \citet{keller_carbon_1993}.}
    \label{fig:composition}
\end{figure}

\subsection{Mass flux}

Most IDP particles entering Earth have masses between 0.1 to 1000~$\mu$g \citep[Figure~\ref{fig:Idp_mass_dist} and][]{cplane_cosmic_2012}. The total mass influx of these small particles is estimated to be larger than that of large meteorites like Chelyabinsk \citep{popova_chelyabinsk_2013} and Tunguska \citep{kulik_results_1940}. However, the mass flux of IDPs reaching Earth is difficult to estimate. Various measurements \citep{wasson_comment_1987, love_direct_1993, peucker-ehrenbrink_accretion_1996, mathews_micrometeoroid_2001, gabrielli_meteoric_2004, voigt_nitric_2005, lanci_meteoric_2006, lanci_meteoric_2007, hervig_first_2009, nesvorny_cometary_2010, rojas_micrometeorite_2021} and theoretical models \citep{cziczo_ablation_2001, plane_time-resolved_2004, vondrak_chemical_2008, gardner_seasonal_2011, drolshagen_mass_2017} have found IDP mass fluxes between 5 to 300 tonnes per day \citep{cplane_cosmic_2012}. This is an uncertainty of two orders of magnitude for the planet that we can observe best. In addition to measurement uncertainties, there are natural variations in the mass flux, for example, from Earth's relative position to Jupiter \citep{scafetta_60-year_2020}. Estimates for Mars and Venus are similar to Earth \citep[$\sim$10 tonnes per day;][]{ceplecha_meteor_1998, frantseva_delivery_2018}. For the outer Solar System, dynamical dust tracing simulations found values from 5 to 600 tonnes per day, depending on their size and distance to the Sun \citep{grun_collisional_1985, poppe_improved_2016}.  Given these uncertainties, we assume Earth experiences an IDP mass flux of 50 tonnes per day, defined here as $M = 10^{-16}~\mathrm{g~cm^{-2}s^{-1}} = 31.54 ~\mathrm{\mu g~m^{-2}yr^{-1}}$.

\begin{table}
    \centering
    \caption{Variables and parameters from Eq.~\ref{eq:cabmod_dvdt} to \ref{eq:pvap_fe}.}
    \label{tab:cabmod_vars}
    \begin{tabular}{l l l l}
        \hline\hline
        Variable                       &                  & units         & default \\  
        \hline 
        Time                           & $t$              & s             & -       \\
        Ideal gas constant             & $R_g$            & $\frac{\mathrm{erg}}{\mathrm{K~mol}}$ & 8.31$\times$$10^{7}$ \\
        Stefan-Boltzmann constant      & $\sigma_\mathrm{SB}$ & $\frac{\mathrm{W}}{\mathrm{m^2K^4}}$ & 5.67$\times$$10^{-8}$ \\
        Boltzmann constant             & $k_b$            & $\frac{\mathrm{erg}}{\mathrm{K}}$ & 1.38$\times$$10^{-16}$ \\
        Atmospheric                    &                  &               &         \\
        $\rightarrow$ pressure         & $p$              & bar    & -       \\
        $\rightarrow$ temperature      & $T_a$            & K             & -       \\
        $\rightarrow$ effective temperature  & $T_\mathrm{eff}$  & K             & 1500    \\
        $\rightarrow$ density          & $\rho_a$         & $\frac{\mathrm{g}}{\mathrm{cm^3}}$      & -       \\
        $\rightarrow$ gravity          & $g$              & $\frac{\mathrm{m}}{\mathrm{s^2}}$       & 316     \\
        $\rightarrow$ mean molecular weight & $\mu$       & $\frac{\mathrm{g}}{\mathrm{mol}}$         & 2.34    \\
        IDP                            &                  &               &         \\
        $\rightarrow$ velocity         & $v$              & $\frac{\mathrm{km}}{\mathrm{s}}$          & -       \\
        $\rightarrow$ entry velocity   & $v_0$            & $\frac{\mathrm{km}}{\mathrm{s}}$          & 30      \\
        $\rightarrow$ total mass       & $m_\mathrm{tot}$ & $\mu$g        & 10      \\
        $\rightarrow$ material mass    & $m_i$            & $\mu$g        & -       \\
        $\rightarrow$ total density    & $\rho_m$         & $\frac{\mathrm{g}}{\mathrm{cm^3}}$     & -       \\
        $\rightarrow$ total radius     & $r$              & cm            & -       \\
        $\rightarrow$ material radius  & $r_i$            & cm            & -       \\
        $\rightarrow$ temperature      & $T$              & K             & -       \\
        $\rightarrow$ entry angle      & $\chi$           & degree        & 30      \\
        $\rightarrow$ vapour pressure  & $p_i^\mathrm{vap}$ & $\frac{\mathrm{dyn}}{\mathrm{cm^2}}$  & -       \\
        Material: MgSiO$_3$            &                  &               &         \\
        $\rightarrow$ specific heat    & $C_i$            & $\frac{\mathrm{J}}{\mathrm{kg~K}}$        & 1000    \\
        $\rightarrow$ molecular weight & $\mu_i$          & $\frac{\mathrm{g}}{\mathrm{mol}}$         & 100.4   \\
        $\rightarrow$ latent heat      & $L_i$            & $\frac{\mathrm{MJ}}{\mathrm{kg}}$         & 6       \\
        $\rightarrow$ density          & $\rho_i$         & $\frac{\mathrm{g}}{\mathrm{cm^3}}$      & 3.2     \\
        $\rightarrow$ mass fraction    & $f_i$            & \%            & 64      \\
        Material: Fe                   &                  &               &         \\
        $\rightarrow$ specific heat    & $C_i$            & $\frac{\mathrm{J}}{\mathrm{kg~K}}$       & 450     \\
        $\rightarrow$ molecular weight & $\mu_i$          & $\frac{\mathrm{g}}{\mathrm{mol}}$         & 55.8    \\
        $\rightarrow$ latent heat      & $L_i$            & $\frac{\mathrm{MJ}}{\mathrm{kg}}$         & 0.25    \\
        $\rightarrow$ density          & $\rho_i$         & $\frac{\mathrm{g}}{\mathrm{cm^3}}$      & 7.9     \\
        $\rightarrow$ mass fraction    & $f_i$            & \%            & 23.3    \\
        \hline
    \end{tabular}
\end{table}

In exoplanet systems, dust is observed through excess luminosity in the infrared \citep{moro-martin_dusty_2013, hughes_debris_2018, ertel_hosts_2020, defrere_hosts_2021, garreau_hosts_2025} which varies significantly between planetary systems. While this allows us to determine column number densities of debris disks, the amount of IDPs is difficult to quantify. Observations of debris disks have shown that dust distributions have complex radial structures \citep[e.g.][]{marino_alma_2026, zawadzki_alma_2026}, depend on the age of the system \citep{su_debris_2006, meyer_evolution_2008}, and can temporarily increase due to collisional events \citep[e.g.][]{kenyon_prospects_2005, bottke_late_2017} or planetary alignments \citep{scafetta_60-year_2020}. In general, young systems ($< 1$~Gyr) are expected to have more dust due to remnant material from planet formation \citep{gaspar_collisional_2013}. During planet formation itself, dust infall is at its maximum, with mass flux estimates ranging from $10^{-6}~\mathrm{g~cm^{-2}s^{-1}}$ to $10^{2}~\mathrm{g~cm^{-2}s^{-1}}$ \citep{machida_gas_2010, johansen_forming_2017, thanathibodee_variable_2020}. However, such large mass fluxes can break the model assumptions of \texttt{Nimbus} and the Sonora Diamondback grid. 

\subsection{Composition}

Solar System chondrites and IDPs are rich in oxygen, magnesium, silicon, and iron (Figure~\ref{fig:composition}). These species are also the main cloud forming materials in hot gas-giant exoplanets and brown dwarfs \citep{helling_dust_2006, helling_sparkling_2019, gao_aerosol_2020}. Several observations of clouds in substellar atmospheres have found evidence of enstatite (MgSiO$_3$), forsterite (Mg$_2$SiO$_4$), and quartz (SiO$_2$) \citep[e.g.][]{suarez_ultracool_2022, grant_jwst-tst_2023, miles_jwst_2023, biller_jwst_2024, inglis_quartz_2024, hoch_silicate_2025, molliere_evidence_2025}. In addition to magnesium-silicates, iron is of special interest for exoplanet observations due to its high opacity. If iron ablates, the additional opacity can weaken the spectral features of gas-phase molecules and magnesium-silicates \citep{hervig_constraints_2017, kiefer_why_2024}. Conversely, iron will act as a CCN for ablated magnesium-silicate materials if it does not ablate. In this scenario, iron is `hidden' and has no effect on observations. Our ablation height simulations (Section~\ref{sec:ablation_height}) indicate that, on average, roughly half of the iron content ablates during entry. For this work, we assume the ablated iron is well mixed with the magnesium silicates. The composition of IDPs is assumed to be the same as an average Solar System chondrite (Figure~\ref{fig:composition}). Species other than enstatite and iron are neglected since we focus on the first order effects of IDPs on observations. The ablation products considered are therefore enstatite (64\% of the total IDP mass) and iron (12\% of the total IDP mass).

\subsection{Ablation height}
\label{sec:ablation_height}
    
\begin{figure*}
    \centering
    \includegraphics[width=\linewidth]{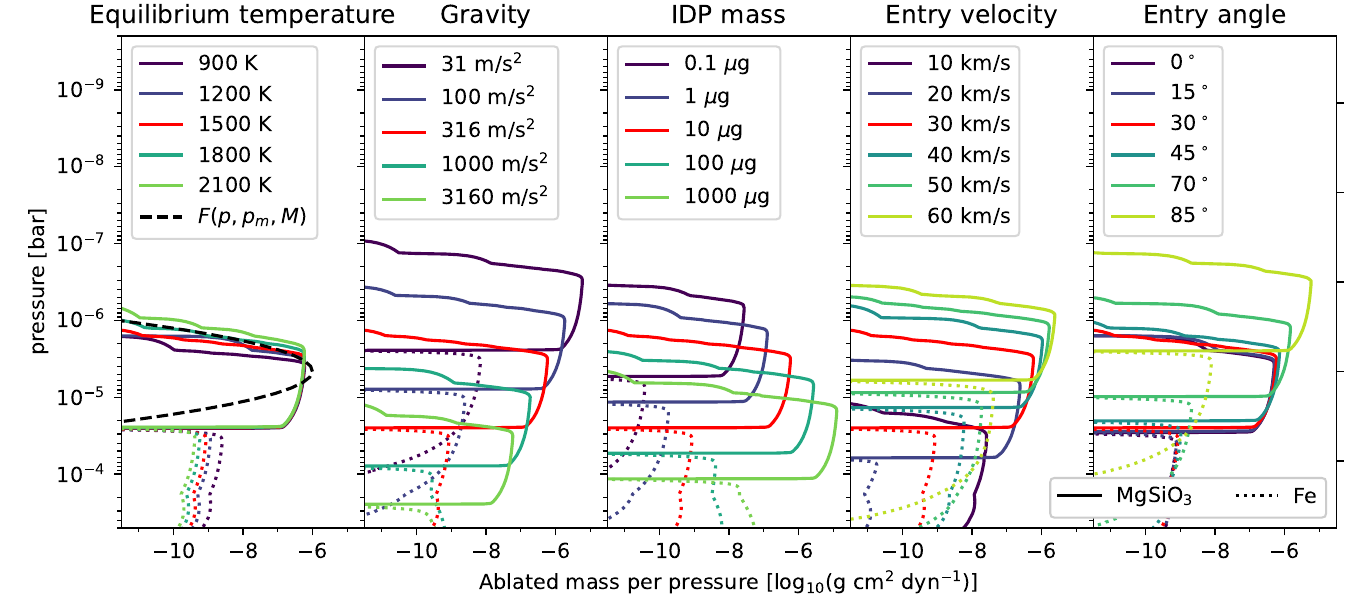}
    \caption{Simulated ablation heights of IDPs in exoplanet atmospheres. Our fiducial simulation is marked in red. Solid and dashed curves signify IDP compositions of MgSiO$_3$ and Fe, respectively. The black dashed curve on the left most panel shows the approximation of the mass injection function $F_v(p)$ for an arbitrary IDP mass flux for comparison with the ablation profiles.}
    \label{fig:abl_height}
\end{figure*}

On Earth, IDPs ablate around 100~km from its surface at a pressure of $\sim10^{-5}$~bar \citep{cziczo_ablation_2001, cplane_cosmic_2012, campbell-brown_high-resolution_2013, jiao_first_2022}. The exact height depends on seasonal effects, local changes in atmospheric pressure, and IDP composition \citep{cplane_cosmic_2012, stober_neutral_2012}. Ablation heights are slightly lower on Mars and higher on Venus due to the thinner and thicker atmosphere, respectively \citep{crismani_detection_2017, carrillo-sanchez_cosmic_2020}. On Solar System gas-giants, ablation heights span a range between 10$^{-4}$ to 10$^{-10}$~bar \citep{moses_dust_2017}. 

Previous theoretical studies suggest that ablation heights for exoplanet gas-giants are comparable to those of our Solar System's gas-giants \citep{arras_dust_2022}. However, the large diversity of exoplanet atmospheres warrants a detailed parameter study to better characterize ablation heights. To achieve this, we simplify the Chemical Ablation Model \citep[CABMOD;][]{vondrak_chemical_2008} and apply it to the Sonora Diamondback grid \citep{morley_sonora_2024}. Our simplified ablation model solves the time evolution of the IDP velocity $v$, mass $m_i$, altitude (in pressure coordinates $p$), and temperature $T$ during entry into the atmosphere:
\begin{align}
    \label{eq:cabmod_dvdt}
    \frac{dv}{dt} &= - v^2 \frac{3 \rho_a}{4\rho_m r} + g \\
    \label{eq:cabmod_dmdt}
    \frac{dm_i}{dt} &= - 4 \pi r_i^2 p_i^\mathrm{vap} \sqrt{\frac{\mu_i}{2 \pi R_g T}} \\
    \label{eq:cabmod_dpdt}
    \frac{dp}{dt} &= - v \cos(\chi) g \rho_a \\
    \label{eq:cabmod_dTdt}
    \frac{dT}{dt} &= \frac{\frac{\pi}{2}r^2 v^3 \rho_a - 4 \pi r^2 \sigma_\mathrm{SB} (T^4 - T_a^4) - \sum_iL_i\frac{dm_i}{dt}}{\sum_i \frac{4}{3} \pi r_i^3 \rho_i C_i} 
\end{align}
The list of variables can be found in Table~\ref{tab:cabmod_vars}. For all Sonora Diamondback models, we assume solar metallicity, solar C/O ratio, and $f_\mathrm{sed} = 2$. All profiles are extended isothermally at the top of the atmosphere. The initial IDP temperature is equal to the isothermal extension. Due to the rapid heating upon entry of the atmosphere, the choice of initial temperature is of little consequence. The radii of the IDP ($r$) and its constituent materials ($r_i$), in conjunction with the atmospheric ($\rho_a$) and IDP ($\rho_m$) mass density, are derived at runtime:
\begin{align}
    r &= \sqrt[3]{\frac{3 m_\mathrm{tot}}{4 \pi \rho_m}} \\
    r_i &= \sqrt[3]{\frac{3 m_i}{4 \pi \rho_i}} \\
    \rho_a &= \frac{p \mu}{T_a R_g} \\
    \rho_m &= \frac{m_\mathrm{tot}}{\sum_i \frac{m_i}{\rho_i}}
\end{align}
The vapour pressures of enstatite and iron are \citep{visscher_atmospheric_2010, morley_sonora_2024}:
\begin{align}
    p^\mathrm{vap}_\mathrm{MgSiO_3} &= 10^{19.43 - 28665/T_e} ~\frac{\mathrm{dyn}}{\mathrm{cm^2}}\\
    \label{eq:pvap_fe}
    p^\mathrm{vap}_\mathrm{Fe} &= 10^{7.09 - 20995/T_e}~\frac{\mathrm{dyn}}{\mathrm{cm^2}}
\end{align}
To determine the ablation height for a range of IDPs and atmospheres, we vary the effective temperature $T_\mathrm{eff}$, gravity $g$, IDP mass $m$, entry velocity $v_0$, and entry angle $\chi$. In the Solar System, typical IDP masses range from 0.1~$\mu$g to 1000~$\mu$g \citep{cplane_cosmic_2012} and typical entry velocities are between 10~km~s$^{-1}$ to 60~km~s$^{-1}$ \citep{le_feuvre_nonuniform_2008, minton_dynamical_2010}. While varying one parameter, all others are fixed to the values listed in Table~\ref{tab:cabmod_vars}.

Our results (Figure~\ref{fig:abl_height}) show that typical ablation heights of IDPs in gas-giant exoplanets are between 10$^{-4}$~bar, for high gravity planets and heavy IDPs, to 10$^{-7}$~bar, for low gravity planets and light IDPs. The effective temperature of the planet and the entry angle of the IDPs are less significant, yielding nearly the same ablation height in most cases. These results are in agreement with the work of \citet{arras_dust_2022}, who simulated a Jupiter-like case with $g=27$~m~s$^{-2}$, and \citet{lavvas_aerosol_2017}, who simulated HD~189733~b.

For IDPs that are fast enough for near complete ablation of enstatite ($v_0 >$10 km~s$^{-1}$) but that are within reasonable limits for the assumptions of our model ($v_0 <$ 100 km~s$^{-1}$), we find an empirical relationship for the pressure of maximum ablation efficiency $p_\mathrm{m}^\mathrm{MgSiO_3}$ [units of bar]:
\begin{align}
    \nonumber \log_{10}&\left( \frac{p_\mathrm{m}^\mathrm{MgSiO_3}}{[\mathrm{bar}]}\right) = \log_{10}\left( \frac{g}{[\mathrm{m~s^{-2}}]}\right) + \\ 
    \label{eq:ablation_height}
    &0.33 \log_{10}\left( \frac{m}{[\mathrm{\mu g}]}\right) - 1.9 \log_{10}\left( \frac{v}{[\mathrm{km~s^{-1}}]}\right) - 5.48
\end{align}
Our results from Figure~\ref{fig:abl_height} show that iron ablates at roughly 10 times the pressure of enstatite. We therefore assume:
\begin{equation}
    p_\mathrm{m}^\mathrm{Fe} = 10 ~p_\mathrm{m}^\mathrm{MgSiO_3}
\end{equation}
To approximate the mass injection function $F_v(p)$~[units of g~cm$^{-3}$s$^{-1}$] from ablated IDPs, we use a log-normal distribution:
\begin{align}
    F_v(p) = \frac{M g \rho_a}{p \sigma \sqrt{2 \pi}} \exp \left(\frac{-(\ln(p) - \ln(p_\mathrm{m}))^2}{2 \sigma^2}\right)
\end{align}
where $\sigma$ is the standard deviation of the log-normal function. We find that $\sigma = 0.5$ provides an adequate fit to the ablation profiles of IDPs (see Figure~\ref{fig:abl_height}).


\section{Cloud modelling}
\label{sec:nimbus}

\subsection{Cloud structures with \texttt{Nimbus}}

In order to assess the effect of IDPs on the cloud structure of exoplanet gas-giants, we use the cloud model \texttt{Nimbus} \citep{kiefer_connecting_2026}. To account for IDPs, four improvements were made:
\begin{enumerate}
    \item IDPs present a source of material. We therefore included a mass injection function for the gas-phase $F_v^\mathrm{mat}$~[units of g~cm$^{-3}$s$^{-1}$], cloud particle number density $F_n^\mathrm{mat}$~[units of g~cm$^{-3}$s$^{-1}$], and cloud particle mass $F_c^\mathrm{mat}$~[units of g~cm$^{-3}$s$^{-1}$].
    \item IDPs add magnesium, silicon, and iron which will all condense onto each other. We therefore add the physics of mixed cloud particles into \texttt{Nimbus}.
    \item The opacities of mixed particles differ from homogeneous materials. We therefore calculate the effective refractive index using the Landau-Lifshitz-Looyenga approximation \citep{landau_electrodynamics_1960, looyenga_dielectric_1965} and the opacities using Mie theory \citep{mie_beitrage_1908}.
    \item Coagulation and gravitational coalescence reduce the number of CCNs provided from high altitude nucleation of IDPs. We therefore implement both into \texttt{Nimbus}. 
\end{enumerate}
The cloud model \texttt{Nimbus} solves the time evolution of the gas-phase MMR $q_v^\mathrm{mat}$, cloud particle MMR $q_c^\mathrm{mat}$, and CCN MMR $q_n$. It considers the formation of new cloud particles via nucleation, growth of cloud particles via condensation, particle and gas-phase transport via diffusion and gravitational settling, and material influx through ablating IDPs: 
\begin{align}
    \label{eq:nimb_3}
    \nonumber \rho_a \frac{d q_v^\mathrm{mat}}{dt} = &-\frac{\partial}{\partial z} K_{zz} \rho_a \frac{\partial q_v^\mathrm{mat}}{\partial z} \\
    &- m_\mathrm{n} J^\mathrm{mat} - m_\mathrm{1}^\mathrm{mat} G^\mathrm{mat}  + F_v^\mathrm{mat} \\
    \label{eq:nimb_1}
    \nonumber \rho_a \frac{d q_c^\mathrm{mat}}{dt} = &\frac{\partial}{\partial z} q_c^\mathrm{mat} \rho_a v_\mathrm{dr} - \frac{\partial}{\partial z} K_{zz} \rho_a \frac{\partial q_c^\mathrm{mat}}{\partial z} \\
    &+ m_\mathrm{n} J^\mathrm{mat} + m_\mathrm{1}^\mathrm{mat} G^\mathrm{mat} + F_c^\mathrm{mat} + F_n^\mathrm{mat}\\
    \label{eq:nimb_2}
    \nonumber \rho_a \frac{d q_n}{dt} = &\frac{\partial}{\partial z} q_n \rho_a v_\mathrm{dr} - \frac{\partial}{\partial z} K_{zz} \rho_a \frac{\partial q_n}{\partial z} - m_n f_\mathrm{coag} \\
    &- m_n f_\mathrm{coal} + \sum_\mathrm{mat} (m_\mathrm{n} J^\mathrm{mat} + F_n^\mathrm{mat})
\end{align}
where $K_{zz}$ [units of cm$^2$s$^{-1}$] is the diffusion constant, $z$~[units of cm] is the altitude, $J^\mathrm{mat}$~[units of cm$^{-3}$s$^{-1}$] is the nucleation rate, and $G^\mathrm{mat}$~[units of cm$^{-3}$s$^{-1}$] is the growth rate, $m_\mathrm{n}$~[units of g] is the CCN mass, $m_1^\mathrm{mat}$~[units of g] is the mass of a single unit of a cloud forming material, and $v_\mathrm{dr}$~[units of cm~s$^{-1}$] is the settling velocity. 

The coagulation $f_\mathrm{coag}$~[Units of cm$^{-3}$s$^{-1}$] and gravitational coalescence rate $f_\mathrm{coal}$~[Units of cm$^{-3}$s$^{-1}$] are calculated following \citet{lee_beyond_2025} for mono-disperse particle distributions:
\begin{align}
    f_\mathrm{coag} &= \frac{4 k_bT \beta}{3 \eta_a} n_c^2\\
    f_\mathrm{coal} &= 2 \pi r_c^2 E\Delta v ~n_c^2 
\end{align}
where $n_c$~[Units of cm$^{-3}]$ is the cloud particle number density, $\eta_a$~[Units of g~cm$^{-1}$s$^{-1}$] is the atmospheric dynamical viscosity, and $\Delta v$~[Units of cm~s$^{-1}$] is the relative velocity of cloud particles. The variables $\beta$ and $E$ are defined as:
\begin{align}
    \beta &= 1 + \mathrm{Kn} \left[1.165 + 0.483 \exp \left(\frac{-0.997}{\mathrm{Kn}} \right) \right]\\
    E &= \begin{cases} \mathrm{Kn} < 1, & \max \left[ 0, 1 - 0.42 (\frac{v_\mathrm{dr} \Delta v}{g r_c})^{-0.75} \right] \\ \mathrm{Kn} \leq 1 & 1\end{cases}
\end{align}
where $\mathrm{Kn} = l/r_c$ is the Knudsen number, $l$~[Units of cm] is the mean free path, and $r_c$~[Units of cm] is the cloud particle radius.

Equations~\ref{eq:nimb_3} and \ref{eq:nimb_1} are solved for each cloud particle material individually. The total cloud particle density ($\rho_c$) and radius ($r_c$) are calculated as:
\begin{align}
    \rho_c &= \frac{\sum_\mathrm{mat} q_c^\mathrm{mat}}{\sum_\mathrm{mat} \frac{q_c^\mathrm{mat}}{\rho_c^\mathrm{mat}}} \\
    r_c &= \sqrt[3]{\frac{3}{4 \pi \rho_c}\frac{m_n\sum_\mathrm{mat} q_c^\mathrm{mat}}{q_n}}
\end{align}
A detailed description of the assumptions and numerics of \texttt{Nimbus} can be found in \citet{kiefer_connecting_2026}.

\subsection{Synthetic spectra calculations}

Optical properties of cloud particles can be complex and depend on their composition \citep{wakeford_transmission_2015}, shape \citep{min_scattering_2003, min_shape_2003, min_absorption_2006, samra_mineral_2020, samra_mineral_2022}, and morphology \citep{kiefer_why_2024, moran_fractal_2025, lodge_fractal_2026}. For this work, we assume well-mixed spherical particles. The refractive indices of mixed materials are calculated using the Landau-Lifshitz-Looyenga method \citep{landau_electrodynamics_1960, looyenga_dielectric_1965}, and the extinction, scattering, and asymmetry coefficients are calculated with Mie-theory \citep{mie_beitrage_1908}. We use MieNet \citep{attaway_mienet_2026} to calculate cloud particle opacities for all wavelengths, radii, and pressures. This software package allows us to pre-calculate Mie-opacity grids which significantly reduces the computation time. The opacity data and radii interpolation scheme are taken from \texttt{Virga} \citep{batalha_natashabatalhavirga_2020, batalha_condensation_2026, moran_fractal_2025, lodge_fractal_2026}.

The cloud particle opacities are passed to the radiative transfer code \texttt{PICASO} \citep{batalha_exoplanet_2019, mukherjee_picaso_2023, mang_picaso_2026} to produce the synthetic transmission and thermal emission spectra. We assume a Sun-like host star, for all transmission spectra. The following gas-phase opacities are considered: 
H$_2$O \citep{polyansky_exomol_2018}, 
CO$_2$ \citep{huang_reliable_2014}, 
CH$_4$ \citep{yurchenko_vibrational_2013, yurchenko_exomol_2014}, 
NH$_3$ \citep{yurchenko_variationally_2011, wilzewski_h2_2016}, 
N$_2$ \citep{rothman_hitran2012_2013}, 
CO \citep{rothman_hitemp_2010, gordon_hitran2016_2017, li_accounting_2015}, 
TiO \citep{mckemmish_exomol_2019, gharib-nezhad_exoplines_2021}, 
VO \citep{mckemmish_exomol_2016, gharib-nezhad_exoplines_2021}, 
and 
FeH \citep{dulick_line_2003, hargreaves_high-resolution_2010}.

\section{Interplanetary Dust Particles in Exoplanet Atmospheres}
\label{sec:obs}

\begin{figure}
    \centering
    \includegraphics[width=\linewidth]{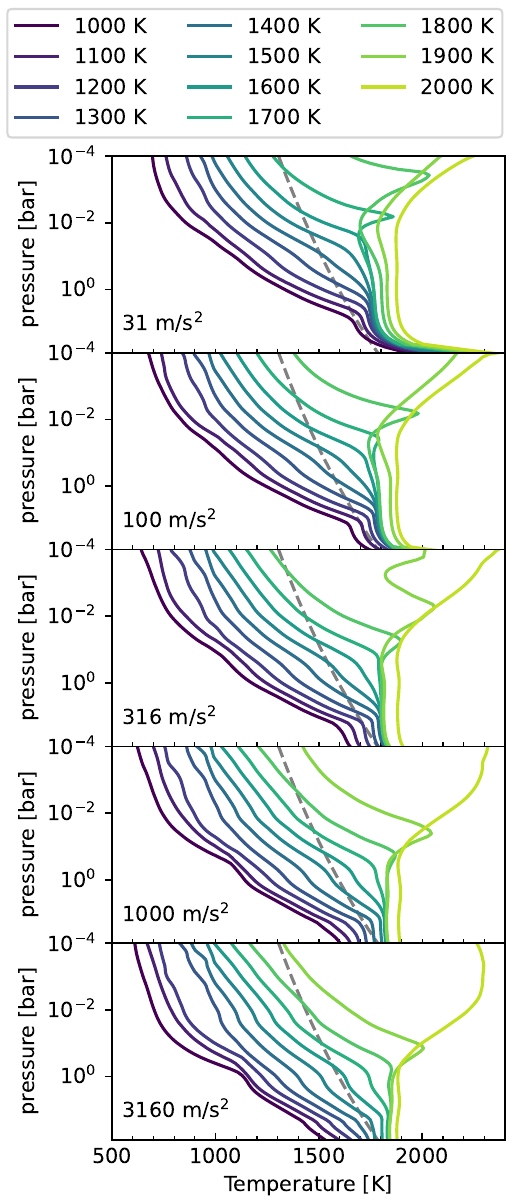}
    \caption{Temperature pressure profiles of irradiated exoplanets at different equilibrium temperatures and surface gravities calculated with Picaso. The gray dashed curve shows the condensation curve of MgSiO$3$.}
    \label{fig:picaso_grid}
\end{figure}

\begin{figure*}
    \centering
    \includegraphics[width=\linewidth]{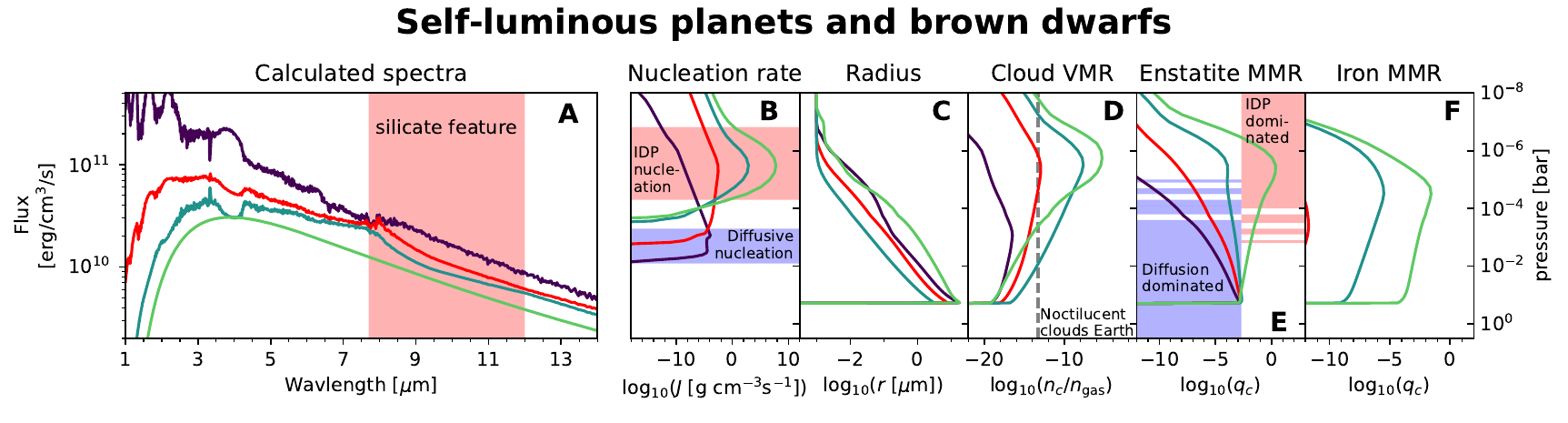}
    \includegraphics[width=\linewidth]{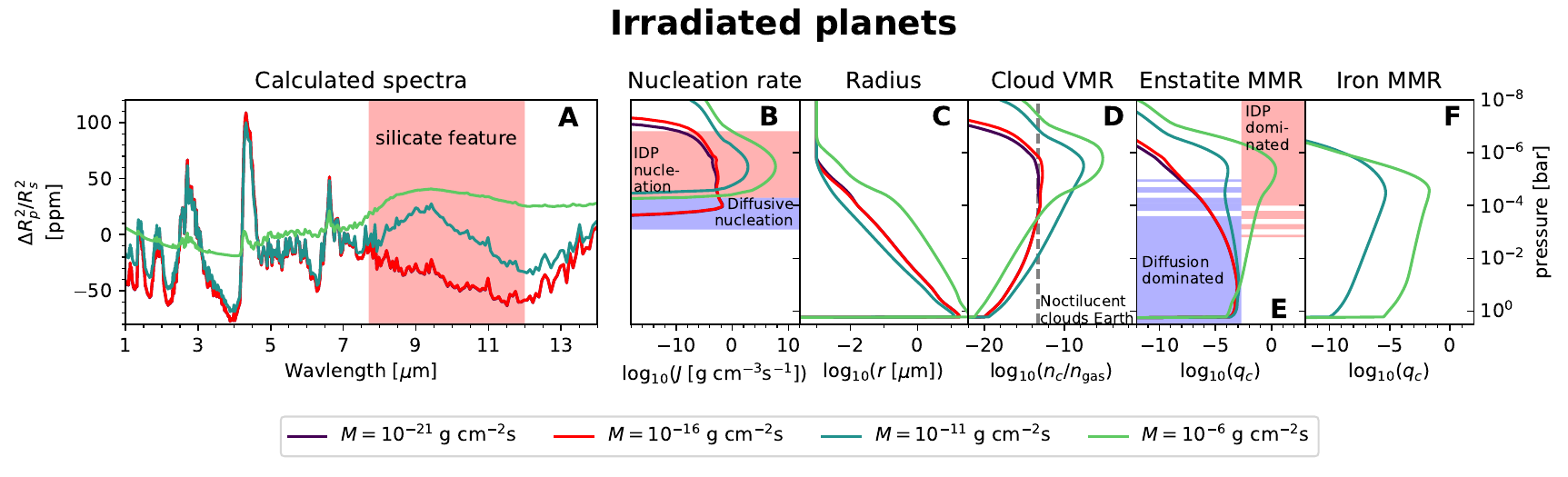}
    \caption{Example cloud structure for $T_\mathrm{eff} = 1500$~K, $g = 316~\mathrm{m~s^{-2}}$, $K_{zz} = 10^{8}~\mathrm{cm^{2}s^{-1}}$, and a range of IDP mass fluxes. The VMR of Earth noctilucent clouds is taken from \citet{baumgarten_particle_2008}. \textbf{Top:} Self-luminous planets and brown dwarfs. \textbf{Bottom:} Irradiated planets. The transmission spectra are offset by the mean transit depth between 4.2 to 7.2 $\mu$m.}
    \label{fig:example_from_grid}
\end{figure*}

\begin{figure*}
    \centering
    \includegraphics[width=\linewidth]{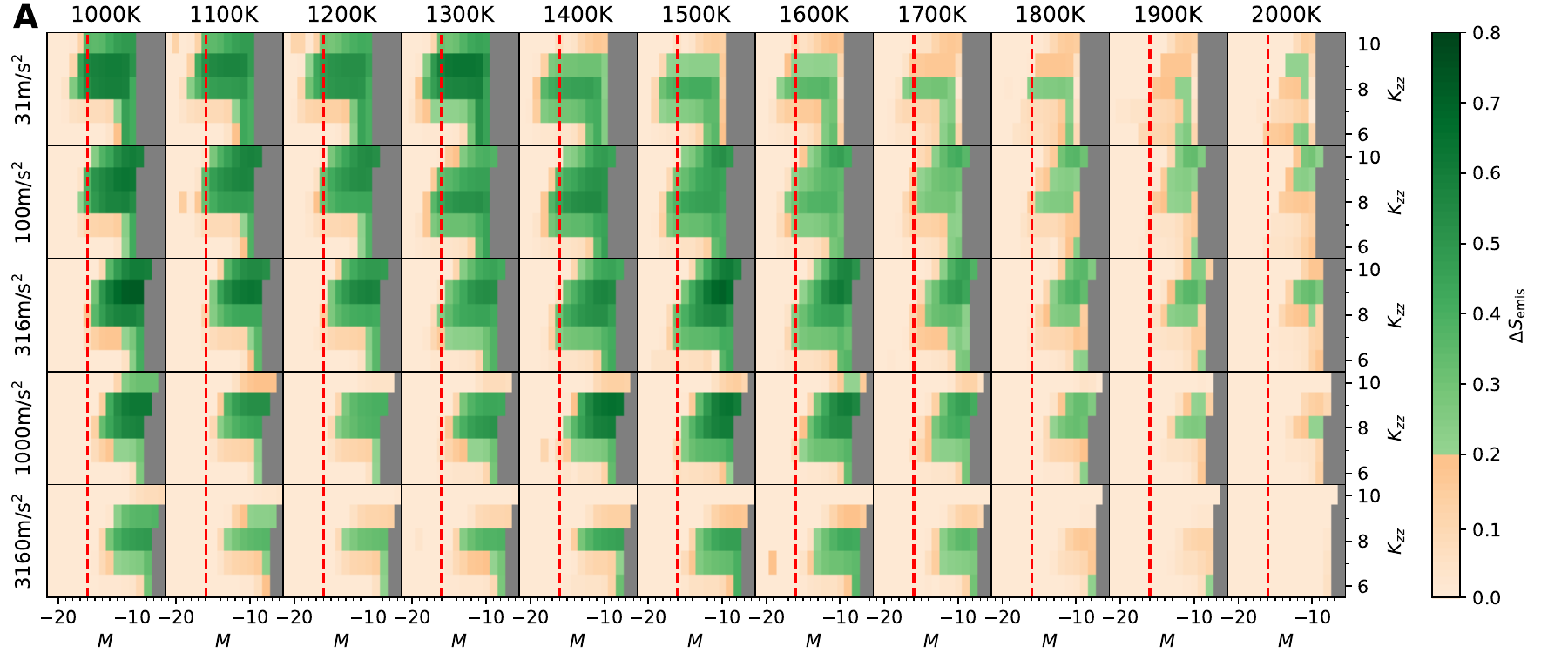}
    \includegraphics[width=\linewidth]{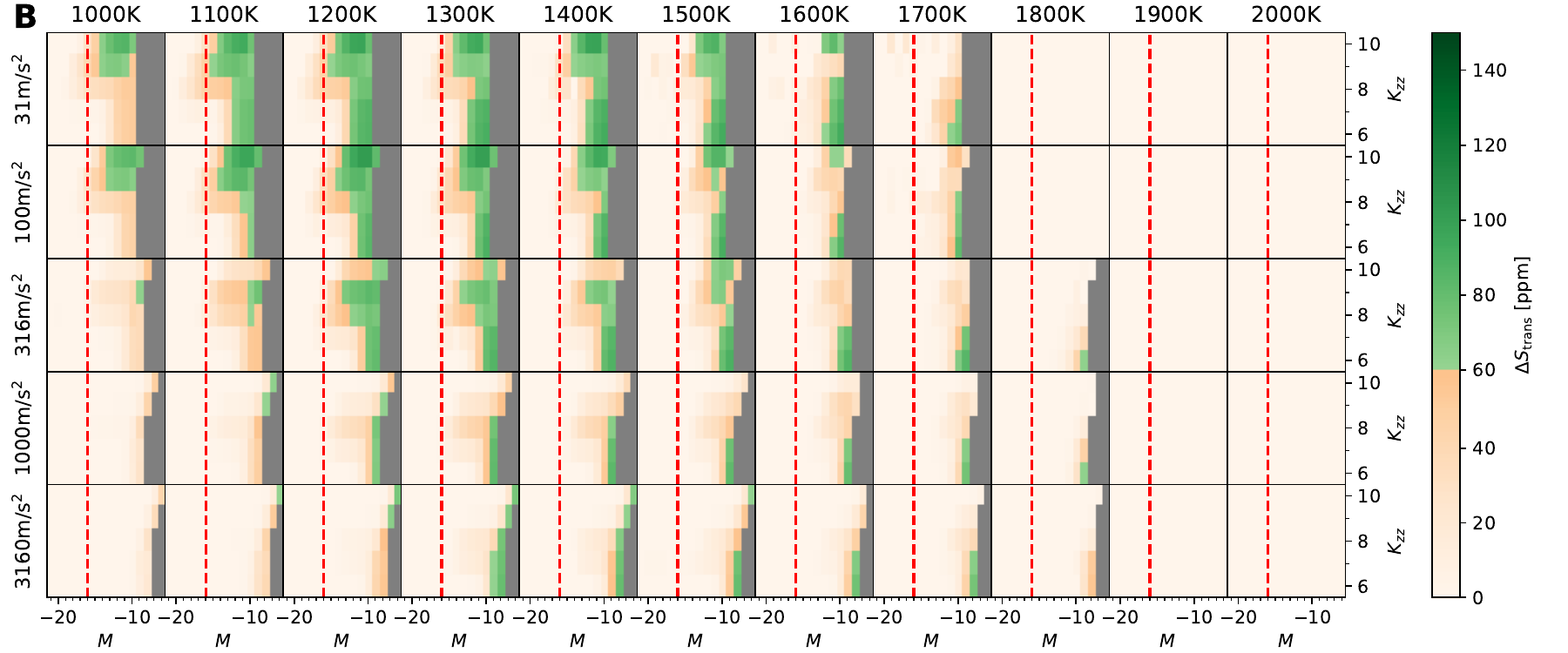}
    \caption{Impact of IDPs on \texttt{Nimbus} cloud structures over the Sonora Diamondback grid. Panel A: Silicon index change in thermal emission spectra. Panel B: Silicon index change in transmission spectra. Gray panels reach enstatite MMRs outside of our model assumptions (MMR$_\mathrm{MgSiO_3} > 1$\%). Orange values are below the typical observable limit for thermal emission ($\Delta S_\mathrm{emis} < 0.2$) or transmission spectra ($\Delta S_\mathrm{trans} < 60~\mathrm{ppm}$). The red dashed curves mark Earth-like IDP mass fluxes.}
    \label{fig:grid}
\end{figure*}

\begin{figure*}
    \centering
    \includegraphics[width=\linewidth]{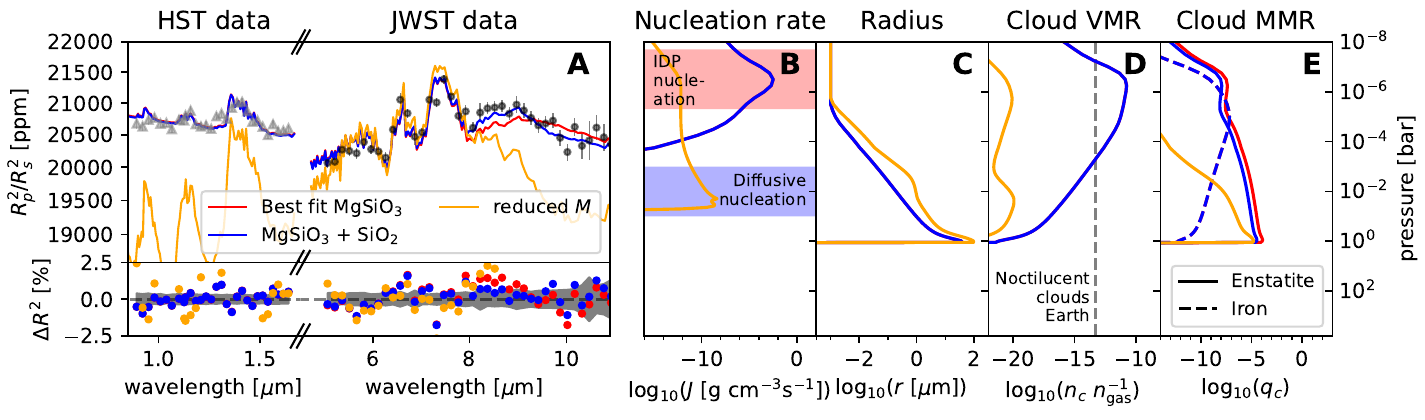}
    \caption{Transmission spectrum and cloud structure of WASP-107~b for enstatite clouds (MgSiO$_3$), mixed enstatite and quartz clouds (MgSiO$_3$ + SiO$_2$), and reduced IDP mass flux of $M = 10^{-21}~\mathrm{g~cm^{-2}s^{-1}}$ (reduced $M$). The VMR of Earth noctilucent clouds is taken from \citet{baumgarten_particle_2008}.}
    \label{fig:wasp107b}
\end{figure*}

\subsection{Model grid}
\label{sec:obs_grid}

To assess the effect of IDPs on the cloud structure of exoplanets and brown dwarfs, we apply \texttt{Nimbus} to all combinations of the following parameter space:
\begin{itemize}
    \item $1000 ~\mathrm{K} \leq T_\mathrm{eff}~\mathrm{or}~T_\mathrm{eq} \leq 2000 ~\mathrm{K}$
    \item $31 ~\mathrm{m~s^{-2}} \leq g \leq 3160 ~\mathrm{m~s^{-2}}$
    \item $10^{6} ~\mathrm{cm^2s^{-1}} \leq K_{zz} \leq 10^{10} ~\mathrm{cm^2s^{-1}}$
    \item $10^{-21} ~\mathrm{g~cm^{-2}s^{-1}} \leq M \leq 10^{-6} ~\mathrm{g~cm^{-2}s^{-1}}$
\end{itemize}
For self-luminous objects, we use the Sonora Diamondback grid \citep{morley_sonora_2024} with $f_\mathrm{sed} = 2$. For irradiated objects, we use \texttt{PICASO} \citep{batalha_exoplanet_2019, mukherjee_picaso_2023, mang_picaso_2026} to calculate the temperature pressure profiles, assuming a sun-like host star (Figure~\ref{fig:picaso_grid}). The orbital distance to the host star is given by the equilibrium temperature $T_\mathrm{eq}$~[Units of K] of the planet, assuming a Bond albedo of 0. In all cases, we assume solar metallicity and solar C/O ratio, an intrinsic temperature of $T_\mathrm{int} = 200$~K, and a heat redistribution factor of 0.5. The ablation height is estimated using Equation~\ref{eq:ablation_height} with $v_0 = 30~\mathrm{km~s^{-1}}$, and $m_\mathrm{tot} = 10~\mathrm{\mu g}$. The example for $T_\mathrm{eff} = 1500$~K, $g = 316~\mathrm{m~s^{-2}}$, and $K_{zz} = 10^{8}~\mathrm{cm^{2}s^{-1}}$ is shown in Figure~\ref{fig:example_from_grid}. To assess the impact of IDPs over the whole parameter space, we use the silicon index. For the thermal emission spectra from self-luminous objects, we follow \citet{suarez_ultracool_2022} and define:
\begin{align}
    S_\mathrm{emis} = \frac{1}{A_9}\left( A_{7.5} + \frac{3}{8}(A_{11.5} - A_{7.5})\right)
\end{align}
For the transmission spectra of irradiated planets, we use:
\begin{align}
    S_\mathrm{trans} = A_9 - \left( A_{6} + \frac{6}{13}(A_{12.5} - A_{6})\right)
\end{align}
where $A_i$ is the average signal in the interval $i \pm 0.3~\mu\mathrm{m}$. To isolate the effect of IDPs, we calculate the difference in silicon index:
\begin{align}
    \Delta S(M) = S(M) - S(10^{-21} ~\mathrm{g~cm^{-2}s^{-1}})
\end{align}
The silicon index differences for thermal emission and transmission spectra are shown in panels A and B of Figure~\ref{fig:grid}, respectively.

The observability of differences in the silicon feature depends strongly on the exoplanet system and is difficult to determine a priori. \citet{suarez_ultracool_2022} have shown that differences of $S_\mathrm{emis} > 0.2$ are often distinguishable with the Spitzer space telescope. For transmission spectra, the transit depth depends on the size of the star. Scaling previous JWST/MIRI observations to a sun-like star, we find typical accuracies reach down to 60~ppm for wavelengths around 9~$\mu$m \citep{madhusudhan_new_2025, holmberg_mid-infrared_2026}, which is also in line with detector limitations \citep{bouwman_spectroscopic_2023}. We use these values as current instrumental limitations when assessing if the effect of IDPs on cloud structures is observable in thermal emission and transmission spectra.

It is important to note, that certain combination of parameters within our model grid might be unlikely to occur. For example, ultra hot Jupiters are at orbital separations where IDPs are likely to evaporate due to radiative heating before reaching the planet. For this work, we take an agnostic approach and calculate the whole parameter space.

\subsection{How do interplanetary dust particles affect exoplanet clouds?}

\subsubsection{Effect on the cloud structure}

The ablation products of IDPs nucleate at high altitudes, providing an additional source of CCNs which increases the total cloud particle number density (Panel D of Figure~\ref{fig:example_from_grid}). Since the cloud forming materials from the interior of the planet now distribute over more particles, the cloud particle radius becomes smaller (Panel C of Figure~\ref{fig:example_from_grid}). These changes lead to a higher cloud particle opacity, strengthening the silicate feature around 9~$\mu$m. For very large IDP mass fluxes, an extreme case is reached where all CCNs are produced from meteoritic material (Panel B of Figure~\ref{fig:example_from_grid}). In this case, coagulation becomes increasingly dominant, leading to an increase in particle size (Panel C of Figure~\ref{fig:example_from_grid}) and a reduction in cloud particle number densities (Panel D of Figure~\ref{fig:example_from_grid}).

Whether nucleation in gas-giant exoplanets occurs bottom-up or top-down is debated. The cloud model \texttt{CARMA} \citep{gao_microphysics_2018, powell_formation_2018, gao_aerosols_2021, komacek_patchy_2022, mang_microphysical_2024, powell_two-dimensional_2024} and previous studies with \texttt{Nimbus} \citep{kiefer_connecting_2026} assume bottom-up nucleation where material is diffusively replenished from the interior of the planet. On the other hand, the cloud model \texttt{DRIFT} \citep{helling_dust_2006, helling_modelling_2013, lee_dynamic_2016, lee_dynamic_2017, helling_sparkling_2019, samra_clouds_2022, helling_exoplanet_2023, kiefer_under_2024} considers top-down nucleation by assuming a relaxation timescale for cloud-forming materials. These cloud structures are characterised by CCNs formed at high altitudes. Recent analysis of WASP-107~b \citep{huang_cloudy_2026} and PSO~J318 \citep{molliere_evidence_2025} found evidence of such high altitude nucleation. Our work shows, that ablation of IDPs can explain the origin of high altitude nucleation and bridge the gap between bottom-up and top-down cloud models (see also Section~\ref{sec:obs_wasp107b}). This is in agreement with previous studies which already suggested IDPs as an origin for high-altitude dust particles \citep{ohno_grain_2021}.

\subsubsection{Which planets are most affected?}

While an Earth-like IDP mass flux of $10^{-16}~\mathrm{g~cm^{-2}s^{-1}}$ can alter the cloud structure of an exoplanet, its effect on transmission and thermal emission spectra is in many cases near the observational limit of current telescopes (Figure~\ref{fig:grid}). Within our model grid, an Earth-like IDP flux is only detectable for certain cases in thermal emission. However, mass fluxes exceeding $10^{-13}~\mathrm{g~cm^{-2}s^{-1}}$ lead to differences in the silicate feature which are observable with current telescopes in many cases. Young planetary systems ($< 1$~Gyr) are therefore the best targets when searching for observational evidence of IDP infall in exoplanet atmospheres since they are more likely to have a higher IDP mass flux \citep{kenyon_prospects_2005, gaspar_collisional_2013}. At even higher IDP mass fluxes $(M > 10^{-10}~\mathrm{g~cm^{-3}s^{-1}})$, IDPs can become the dominant cloud material source, surpassing the cloud material reservoir of the planet itself (enstatite MMR in Figure~\ref{fig:example_from_grid}). Such high IDP mass fluxes, however, would lead to changes in the planetary composition and are therefore only realistic during planet formation \citep{machida_gas_2010, johansen_forming_2017, thanathibodee_variable_2020}. Since enstatite MMRs above 1\% are outside the reasonable limits of our model assumptions, we do not consider these simulations for our analysis and mark them gray in Figure~\ref{fig:grid}.
 
The extent to which transmission and thermal emission spectra are affected by IDPs depends on the atmospheric conditions. In general, we find that: 
\begin{itemize}
    \item high atmospheric mixing rates ($K_{zz}$) efficiently distribute small particles throught the atmosphere,
    \item colder planets have larger distance between the cloud deck and the IDP ablation height, and
    \item lower gravity planets have more extended atmospheres, increasing the signal of all atmospheric features, including the silicon index. 
\end{itemize}
Observations of high mixing, low temperature, and low gravity planets are therefore the most sensitive to IDP mass fluxes.

In transmission spectra, the largest change in the silicate feature occurs in low temperature, low gravity planets. This is due to the larger distance to the cloud deck and the larger scale height of the planet, respectively.

In thermal emission spectra, a silicate feature occurs only if the silicate particles are at the right distance to the photosphere. If these particles are too low, they become optically thick and emit like a blackbody. If too high, they become optically thin and do not affect the spectrum. We therefore see a correlated temperature and gravity dependence in the strength of the silicate feature, since these parameters determine the distance between the IDP ablation height and the cloud deck. At very high IDP mass fluxes, the IDP ablation products themselves can start to become optically thick. This increases the altitude of the photosphere and reduces the strength of the silicate feature.

\subsection{Case Study: WASP-107~b}
\label{sec:obs_wasp107b}

There are few detailed observations of clouds in transmission spectra of exoplanets \citep{grant_jwst-tst_2023, dyrek_so2_2023, welbanks_high_2024}. The silicate clouds detected in WASP-107~b \citep{anderson_discoveries_2017, spake_helium_2018, kreidberg_water_2018, dyrek_so2_2023, welbanks_high_2024, sing_warm_2024, murphy_evidence_2024} are of special interest for this study, because of the planet's low equilibrium temperature \citep[$T_\mathrm{eq} = 770$~K;][]{murphy_evidence_2024} and low gravity \citep[$g=4.67$~m~s$^{-1}$;][]{anderson_discoveries_2017}. At this temperature, silicate clouds are expected to form below the pressure levels probed by transmission spectra. This raises the question as to why silicate particles were observed at high altitudes when they should be cold-trapped deeper in the atmosphere \citep{parmentier_3d_2013, pelletier_vanadium_2023}. IDPs present a possible answer to this puzzle, as they are a source of silicates and ablate at high altitudes.

To investigate whether the silicate signatures in WASP-107~b are the product of ablated IDPs, we apply \texttt{Nimbus} to the temperature pressure profile of \citet{dyrek_so2_2023} to find the best fit to the observational data presented in \citet{welbanks_high_2024}. We find that $K_{zz} = 10^{7.5}~\mathrm{cm}^2\mathrm{s}^{-1}$ 
and $M = 10^{-14}~\mathrm{g~cm^{-2}s^{-1}}$ can reproduce the observed silicate feature with a reduced $\chi^2 = 3.06$ (Figure~\ref{fig:wasp107b}). This $K_{zz}$ values is roughly 1 order of magnitude smaller than what \citet{welbanks_high_2024} found. For comparison, we also show a model with the lowest IDP mass flux in our grid ($M = 10^{-21}~\mathrm{g~cm^{-2}s^{-1}}$) which does not produce the 9~$\mu$m silicate feature and has insufficient molecular muting in the HST wavelengths.

While our model can produce the 9~$\mu$m silicate feature, there is a mismatch in the shape of the feature since we assumed the clouds to be made of enstatite only instead of mixture of silicates \citep[see][]{dyrek_so2_2023}. The composition of Solar System IDPs (Figure~\ref{fig:composition}) also supports a mixture of material since there is 28\% more silicon than magnesium and not all silicon can be bound in enstatite. To test if a mixture of both materials would result in a better fit, we assume that 28\% of the silicon that would have formed enstatite forms quartz instead. The resulting transmission spectrum (Figure~\ref{fig:wasp107b}) provides a better fit to the observations ($\chi^2 = 2.42$). A mixture of enstatite and quartz is not unexpected in hot gas-giants as SiO is known to efficiently nucleate into SiO$_2$ \citep{lee_dust_2015, bromley_under_2016}. However, if SiO$_2$ only nucleated it would be coated by enstatite and therefore not be visible in observations \citep{kiefer_why_2024}. Instead, enstatite and quartz seem to be well mixed throughout the cloud particle. Assessing the mechanism of mixed SiO$_2$ and MgSiO$_3$ formation requires a kinetic nucleation framework \citep[e.g.][]{boulangier_devloping_2019, kohn_dust_2021, kiefer_fully_2024}. Future studies should explore such effects. 

The spectral feature of WASP-107~b can also be explained by condensate clouds alone. However, this requires clouds at altitudes higher than the main cloud deck \citep{changeat_cloud_2025}, strong vertical mixing combined with parametrised nucleation rates \citep{huang_cloudy_2026}, or very low sticking coefficients \citep{kiefer_connecting_2026}. In this work, we have shown that an Earth-like flux of IDPs to WASP-107~b naturally leads to silicate particles at high altitudes, consistent with observations.

\subsection{Additional evidence for IDP particles in Exoplanet atmospheres}

We have shown that IDPs can cause an observable change in the cloud structure of exoplanet atmosphere and that they can explain the occurrence of small high-altitude magnesium-silicate particles in colder planets. However, cloud structures in exoplanet atmospheres depend on many factors like the atmospheric mixing strength \citep{christie_impact_2021, campos_estrada_msg_2025}, atmospheric dynamics \citep{parmentier_3d_2013, lee_dynamically_2024, tan_large-amplitude_2025}, and the microphysics of cloud formation \citep{ohno_clouds_2020, samra_mineral_2022, kiefer_connecting_2026}. To observationally confirm the presence of IDP particles in exoplanet atmospheres, additional evidence is required.

\citet{ohno_super-rayleigh_2020} have shown that an increasing atmospheric opacity with altitude can lead to a super-Rayleigh slope. While bottom-up cloud formation exclusively leads to decreasing opacity profiles \citep{gao_microphysics_2018, batalha_condensation_2026}, IDP sourced clouds can have decreasing opacity profiles due to high altitude nucleation, leading to a large opacity at high altitudes, and subsequent coagulation, which reduces the cloud particle number density and their opacity throughout the observable part of the atmosphere. Similarly, photochemical hazes can lead to a super-Rayleigh slope as well \citep{ohno_super-rayleigh_2020}. Observations of the Rayleigh slope at visible wavelengths in planets where hazes are not expected could therefore provide additional evidence for the ablation of IDP particles.

Further evidence of IDP ablation might be found through the presence of refractory elements, such as Fe, Mg, and Si, and their molecules. Without an external source, the high-altitude abundance of these elements is reduced because of condensation in deeper layers \citep[cold trapping;][]{parmentier_3d_2013, pelletier_vanadium_2023}. In addition to molecules like, for example, FeO and NiO, which naturally arise through the enrichment of the atmosphere with refractory elements \citep{plane_atmospheric_2003, daly_meteoric_2020}, the heat generated during the IDP's entry allow the formation of otherwise inhibited molecules \citep{rietmeijer_interrelationships_2000, court_meteorite_2009, berezhnoy_formation_2010, plane_impacts_2017, niculescu_production_2020}. However, many of the refractory molecules that form are likely able to condense \citep{helling_sparkling_2019}, which removes them from the gas-phase and prevents their detection. Observations aiming to detect gaseous IDP ablation products should therefore target low pressure layers where condensation is inefficient.

\section{Conclusions}

We have modelled the effect of IDP ablation on the cloud structure of hydrogen-dominated gas-giant exoplanets and brown dwarfs to explore their influence on transmission and thermal emission spectra. Our results show that ablation height is strongly dependent on the planet's gravity and IDP composition, mass, and entry velocity. Ablating IDPs also provide an additional source of CCNs to exoplanet atmospheres, resulting in an increase in cloud particle number density and total cloud mass, and a decrease in the cloud particle radius. All three changes serve to increase the observability of the ~$9$ $\mathrm{\mu m}$ silicate feature of silicate-bearing cloud particles. We also find that an Earth-like IDP mass flux results in changes to our modelled spectra that are, in many cases, at the observational limit of JWST.

We infer from our models that exoplanets with low effective temperatures, high atmospheric mixing, and low gravity, such as WASP-107~b, are most affected by IDPs, making them ideal targets for studying the effects of IDP infall. We have shown that a 100 times Earth-like mass flux can explain the observed high-altitude silicate features of WASP-107~b. The best fit to observations was achieved with a mixture of quartz (SiO$_2$), enstatite (MgSiO$_3$), and iron (Fe), consistent with elemental abundances of Solar System IDPs. 

Ultimately, our results show that IDPs have a measurable effect on the atmospheric structure of gas-giant exoplanets. Observations that target low-temperature, low gravity, and low atmospheric mixing planets must consider such effects. Smaller IDP mass fluxes will likely be observable with the increased sensitivity of future telescopes such as Habitable Worlds Observatory and the Extremely Large Telescope, opening up a new avenue to study exoplanet cloud structures and the dusty environments within exoplanetary systems.

\begin{acknowledgments}
The authors thank Irina Kempf for assistance with proofreading and language editing. This work is supported through a Research Award in Planetary Habitability from the UT Center for Planetary Systems Habitability (CPSH). This is University of Texas Center for Planetary Systems Habitability (CPSH) contribution \#0092. G.J.G is supported by UTIG Postdoctoral Fellowship. Support for program JWST-GO-05474.006-A was provided by NASA through a grant from the Space Telescope Science Institute, which is operated by the Association of Universities for Research in Astronomy, Incorporated, under NASA contract NAS5-26555.
\end{acknowledgments}

\begin{contribution}

S.K. came up with the initial research concept, conducted the simulations, and wrote the manuscript. G.J.G, C.V.M. and S.P.S.G contributed through regular discussions and edited the manuscript.


\end{contribution}

%

\software{numpy \citep{harris_array_2020},  
          matplotlib \citep{hunter_matplotlib_2007}, 
          xarray \citep{hoyer_xarray_2017, hoyer_xarray_2025},
          SciPy \citep{virtanen_scipy_2020},
          Astropy \citep{astropy_collaboration_astropy_2013, astropy_collaboration_astropy_2018, collaboration_astropy_2022},
          miepython \citep{prahl_miepython_2023},
          MieNet \citep{attaway_mienet_2026},
          Virga \citep{batalha_condensation_2026, moran_fractal_2025},
          PICASO \citep{batalha_exoplanet_2019, mukherjee_picaso_2023, mang_picaso_2026},
          Spellchecking and translation tools (\href{https://www.linguee.com/}{Linguee})
          }


\bibliography{references}{}

@article{holmberg_mid-infrared_2026,
	title = {The {Mid}-infrared {Transmission} {Spectrum} of the {Temperate} {Sub}-{Neptune} {TOI}-270 d},
	volume = {1005},
	issn = {2041-8205},
	url = {https://doi.org/10.3847/2041-8213/ae6640},
	doi = {10.3847/2041-8213/ae6640},
	language = {en},
	number = {2},
	urldate = {2026-08-15},
	journal = {The Astrophysical Journal Letters},
	publisher = {The American Astronomical Society},
	author = {Holmberg, Måns and Madhusudhan, Nikku and Binet, Martin and Sarkar, Subhajit and Rigby, Frances E. and Moses, Julianne I.},
	month = jul,
	year = {2026},
	pages = {L53},
}

@article{madhusudhan_new_2025,
	title = {New {Constraints} on {DMS} and {DMDS} in the {Atmosphere} of {K2}-18 b from {JWST} {MIRI}},
	volume = {983},
	issn = {2041-8205},
	url = {https://doi.org/10.3847/2041-8213/adc1c8},
	doi = {10.3847/2041-8213/adc1c8},
	language = {en},
	number = {2},
	urldate = {2026-08-15},
	journal = {The Astrophysical Journal Letters},
	publisher = {The American Astronomical Society},
	author = {Madhusudhan, Nikku and Constantinou, Savvas and Holmberg, Måns and Sarkar, Subhajit and Piette, Anjali A. A. and Moses, Julianne I.},
	month = apr,
	year = {2025},
	pages = {L40},
}

@article{bouwman_spectroscopic_2023,
	title = {Spectroscopic {Time} {Series} {Performance} of the {Mid}-infrared {Instrument} on the {JWST}},
	volume = {135},
	issn = {1538-3873},
	url = {https://doi.org/10.1088/1538-3873/acbc49},
	doi = {10.1088/1538-3873/acbc49},
	language = {en},
	number = {1045},
	urldate = {2026-08-15},
	journal = {Publications of the Astronomical Society of the Pacific},
	publisher = {The Astronomical Society of the Pacific},
	author = {Bouwman, Jeroen and Kendrew, Sarah and Greene, Thomas P. and Bell, Taylor J. and Lagage, Pierre-Olivier and Schreiber, Jürgen and Dicken, Daniel and Sloan, G. C. and Espinoza, Néstor and Scheithauer, Silvia and Coulais, Alain and Fox, Ori D. and Gastaud, René and Glauser, Adrian M. and Jones, Olivia C. and Labiano, Alvaro and Lahuis, Fred and Morrison, Jane E. and Murray, Katherine and Mueller, Michael and Nayak, Omnarayani and Wright, Gillian S. and Glasse, Alistair and Rieke, George},
	month = mar,
	year = {2023},
	pages = {038002},
}

@misc{wyatt_theory_2025,
	title = {Theory of {Exozodi} {Sources} and {Dust} {Evolution}},
	url = {http://arxiv.org/abs/2508.11754},
	doi = {10.48550/arXiv.2508.11754},
	urldate = {2026-08-13},
	publisher = {arXiv},
	author = {Wyatt, Mark C. and Pearce, Tim D. and Pawellek, Nicole and Dodson-Robinson, Sarah and Faramaz-Gorka, Virginie C. and Rebollido, Isabel and Rigley, Jessica K. and Stark, Christopher C.},
	month = aug,
	year = {2025},
	note = {arXiv:2508.11754 [astro-ph.EP]},
}

@misc{currie_exozodi_2026,
	title = {The exozodi spectral effect: {Residual} habitable zone dust may bias {exoEarth} characterization},
	shorttitle = {The exozodi spectral effect},
	url = {http://arxiv.org/abs/2607.14329},
	doi = {10.48550/arXiv.2607.14329},
	urldate = {2026-08-07},
	publisher = {arXiv},
	author = {Currie, Miles H. and Stark, Christopher C. and Alei, Eleonora and Roberge, Aki},
	month = jul,
	year = {2026},
	note = {arXiv:2607.14329 [astro-ph.EP]},
}

@article{prahl_miepython_2023,
	title = {miepython: {Pure} python implementation of {Mie} scattering},
	shorttitle = {miepython},
	url = {https://zenodo.org/records/8218010},
	doi = {10.5281/zenodo.8218010},
	language = {en},
	urldate = {2023-11-16},
	publisher = {Zenodo},
	author = {Prahl, Scott},
	month = aug,
	year = {2023},
	note = {10.5281/zenodo.8218010},
}

@incollection{moro-martin_dusty_2013,
	title = {Dusty {Planetary} {Systems}},
	isbn = {978-94-007-5606-9},
	url = {https://link.springer.com/rwe/10.1007/978-94-007-5606-9_9},
	doi = {10.1007/978-94-007-5606-9_9},
	language = {en},
	urldate = {2026-04-08},
	booktitle = {Planets, {Stars} and {Stellar} {Systems}},
	publisher = {Springer, Dordrecht},
	author = {Moro-Martın, Amaya},
	year = {2013},
	pages = {431--487},
}

@incollection{flynn_extraterrestrial_2002,
	title = {Extraterrestrial {Dust} in the {Near}-{Earth} {Environment}},
	url = {https://ui.adsabs.harvard.edu/abs/2002mea..book...77F},
	urldate = {2026-05-26},
	booktitle = {Meteors in the {Earth}'s {Atmosphere}},
	publisher = {Cambridge University Press},
	author = {Flynn, George J.},
	month = jan,
	year = {2002},
	note = {ADS Bibcode: 2002mea..book...77F},
	pages = {77},
}

@article{astropy_collaboration_astropy_2013,
	title = {Astropy: {A} community {Python} package for astronomy},
	volume = {558},
	issn = {0004-6361},
	shorttitle = {Astropy},
	url = {https://ui.adsabs.harvard.edu/abs/2013A&A...558A..33A},
	doi = {10.1051/0004-6361/201322068},
	urldate = {2026-06-26},
	journal = {Astronomy and Astrophysics},
	publisher = {EDP},
	author = {{Astropy Collaboration} and Robitaille, Thomas P. and Tollerud, Erik J. and Greenfield, Perry and Droettboom, Michael and Bray, Erik and Aldcroft, Tom and Davis, Matt and Ginsburg, Adam and Price-Whelan, Adrian M. and Kerzendorf, Wolfgang E. and Conley, Alexander and Crighton, Neil and Barbary, Kyle and Muna, Demitri and Ferguson, Henry and Grollier, Frédéric and Parikh, Madhura M. and Nair, Prasanth H. and Unther, Hans M. and Deil, Christoph and Woillez, Julien and Conseil, Simon and Kramer, Roban and Turner, James E. H. and Singer, Leo and Fox, Ryan and Weaver, Benjamin A. and Zabalza, Victor and Edwards, Zachary I. and Azalee Bostroem, K. and Burke, D. J. and Casey, Andrew R. and Crawford, Steven M. and Dencheva, Nadia and Ely, Justin and Jenness, Tim and Labrie, Kathleen and Lim, Pey Lian and Pierfederici, Francesco and Pontzen, Andrew and Ptak, Andy and Refsdal, Brian and Servillat, Mathieu and Streicher, Ole},
	month = oct,
	year = {2013},
	note = {ADS Bibcode: 2013A\&A...558A..33A},
	pages = {A33},
}

@article{plane_impacts_2017,
	title = {Impacts of {Cosmic} {Dust} on {Planetary} {Atmospheres} and {Surfaces}},
	volume = {214},
	issn = {1572-9672},
	url = {https://doi.org/10.1007/s11214-017-0458-1},
	doi = {10.1007/s11214-017-0458-1},
	language = {en},
	number = {1},
	urldate = {2025-10-12},
	journal = {Space Science Reviews},
	author = {Plane, John M. C. and Flynn, George J. and Määttänen, Anni and Moores, John E. and Poppe, Andrew R. and Carrillo-Sanchez, Juan Diego and Listowski, Constantino},
	month = dec,
	year = {2017},
	pages = {23},
}

@article{love_direct_1993,
	title = {A {Direct} {Measurement} of the {Terrestrial} {Mass} {Accretion} {Rate} of {Cosmic} {Dust}},
	volume = {262},
	url = {https://www.science.org/doi/10.1126/science.262.5133.550},
	doi = {10.1126/science.262.5133.550},
	number = {5133},
	urldate = {2026-04-28},
	journal = {Science},
	publisher = {American Association for the Advancement of Science},
	author = {Love, S. G. and Brownlee, D. E.},
	month = oct,
	year = {1993},
	pages = {550--553},
}

@article{collaboration_astropy_2022,
	title = {The {Astropy} {Project}: {Sustaining} and {Growing} a {Community}-oriented {Open}-source {Project} and the {Latest} {Major} {Release} (v5.0) of the {Core} {Package}*},
	volume = {935},
	issn = {0004-637X},
	shorttitle = {The {Astropy} {Project}},
	url = {https://doi.org/10.3847/1538-4357/ac7c74},
	doi = {10.3847/1538-4357/ac7c74},
	language = {en},
	number = {2},
	urldate = {2026-04-01},
	journal = {The Astrophysical Journal},
	publisher = {The American Astronomical Society},
	author = {Collaboration, The Astropy and Price-Whelan, Adrian M. and Lim, Pey Lian and Earl, Nicholas and Starkman, Nathaniel and Bradley, Larry and Shupe, David L. and Patil, Aarya A. and Corrales, Lia and Brasseur, C. E. and Nöthe, Maximilian and Donath, Axel and Tollerud, Erik and Morris, Brett M. and Ginsburg, Adam and Vaher, Eero and Weaver, Benjamin A. and Tocknell, James and Jamieson, William and van Kerkwijk, Marten H. and Robitaille, Thomas P. and Merry, Bruce and Bachetti, Matteo and Günther, H. Moritz and Authors, Paper and Aldcroft, Thomas L. and Alvarado-Montes, Jaime A. and Archibald, Anne M. and Bódi, Attila and Bapat, Shreyas and Barentsen, Geert and Bazán, Juanjo and Biswas, Manish and Boquien, Médéric and Burke, D. J. and Cara, Daria and Cara, Mihai and Conroy, Kyle E and Conseil, Simon and Craig, Matthew W. and Cross, Robert M. and Cruz, Kelle L. and D’Eugenio, Francesco and Dencheva, Nadia and Devillepoix, Hadrien A. R. and Dietrich, Jörg P. and Eigenbrot, Arthur Davis and Erben, Thomas and Ferreira, Leonardo and Foreman-Mackey, Daniel and Fox, Ryan and Freij, Nabil and Garg, Suyog and Geda, Robel and Glattly, Lauren and Gondhalekar, Yash and Gordon, Karl D. and Grant, David and Greenfield, Perry and Groener, Austen M. and Guest, Steve and Gurovich, Sebastian and Handberg, Rasmus and Hart, Akeem and Hatfield-Dodds, Zac and Homeier, Derek and Hosseinzadeh, Griffin and Jenness, Tim and Jones, Craig K. and Joseph, Prajwel and Kalmbach, J. Bryce and Karamehmetoglu, Emir and Kałuszyński, Mikołaj and Kelley, Michael S. P. and Kern, Nicholas and Kerzendorf, Wolfgang E. and Koch, Eric W. and Kulumani, Shankar and Lee, Antony and Ly, Chun and Ma, Zhiyuan and MacBride, Conor and Maljaars, Jakob M. and Muna, Demitri and Murphy, N. A. and Norman, Henrik and O’Steen, Richard and Oman, Kyle A. and Pacifici, Camilla and Pascual, Sergio and Pascual-Granado, J. and Patil, Rohit R. and Perren, Gabriel I and Pickering, Timothy E. and Rastogi, Tanuj and Roulston, Benjamin R. and Ryan, Daniel F and Rykoff, Eli S. and Sabater, Jose and Sakurikar, Parikshit and Salgado, Jesús and Sanghi, Aniket and Saunders, Nicholas and Savchenko, Volodymyr and Schwardt, Ludwig and Seifert-Eckert, Michael and Shih, Albert Y. and Jain, Anany Shrey and Shukla, Gyanendra and Sick, Jonathan and Simpson, Chris and Singanamalla, Sudheesh and Singer, Leo P. and Singhal, Jaladh and Sinha, Manodeep and Sipőcz, Brigitta M. and Spitler, Lee R. and Stansby, David and Streicher, Ole and Šumak, Jani and Swinbank, John D. and Taranu, Dan S. and Tewary, Nikita and Tremblay, Grant R. and Val-Borro, Miguel de and Van Kooten, Samuel J. and Vasović, Zlatan and Verma, Shresth and de Miranda Cardoso, José Vinícius and Williams, Peter K. G. and Wilson, Tom J. and Winkel, Benjamin and Wood-Vasey, W. M. and Xue, Rui and Yoachim, Peter and Zhang, Chen and Zonca, Andrea and Contributors, Astropy Project},
	month = aug,
	year = {2022},
	pages = {167},
}

@article{berezhnoy_formation_2010,
	title = {Formation of molecules in bright meteors},
	volume = {210},
	issn = {0019-1035},
	url = {https://www.sciencedirect.com/science/article/pii/S0019103510002587},
	doi = {10.1016/j.icarus.2010.06.036},
	number = {1},
	urldate = {2026-08-04},
	journal = {Icarus},
	author = {Berezhnoy, Alexey A. and Borovička, Jiří},
	month = nov,
	year = {2010},
	pages = {150--157},
}

@article{court_meteorite_2009,
	title = {Meteorite ablation products and their contribution to the atmospheres of terrestrial planets: {An} experimental study using pyrolysis-{FTIR}},
	volume = {73},
	issn = {0016-7037},
	shorttitle = {Meteorite ablation products and their contribution to the atmospheres of terrestrial planets},
	url = {https://www.sciencedirect.com/science/article/pii/S0016703709001483},
	doi = {10.1016/j.gca.2009.03.006},
	number = {11},
	urldate = {2026-08-04},
	journal = {Geochimica et Cosmochimica Acta},
	author = {Court, Richard W. and Sephton, Mark A.},
	month = jun,
	year = {2009},
	pages = {3512--3521},
}

@article{niculescu_production_2020,
	title = {Production of nitric oxide by a fragmenting bolide: {An} exploratory numerical study},
	volume = {43},
	copyright = {© 2020 John Wiley \& Sons, Ltd.},
	issn = {1099-1476},
	shorttitle = {Production of nitric oxide by a fragmenting bolide},
	url = {https://onlinelibrary.wiley.com/doi/abs/10.1002/mma.6205},
	doi = {10.1002/mma.6205},
	language = {en},
	number = {13},
	urldate = {2026-08-04},
	journal = {Mathematical Methods in the Applied Sciences},
	author = {Niculescu, Mihai L. and Silber, Elizabeth A. and Silber, Reynold E.},
	year = {2020},
	pages = {7758--7773},
}

@article{rietmeijer_interrelationships_2000,
	title = {Interrelationships among meteoric metals, meteors, interplanetary dust, micrometeorites, and meteorites},
	volume = {35},
	copyright = {2000 The Meteoritical Society},
	issn = {1945-5100},
	url = {https://onlinelibrary.wiley.com/doi/abs/10.1111/j.1945-5100.2000.tb01490.x},
	doi = {10.1111/j.1945-5100.2000.tb01490.x},
	language = {en},
	number = {5},
	urldate = {2026-08-04},
	journal = {Meteoritics \& Planetary Science},
	author = {Rietmeijer, Frans J. M.},
	year = {2000},
	pages = {1025--1041},
}

@article{defrere_hosts_2021,
	title = {The {HOSTS} {Survey}: {Evidence} for an {Extended} {Dust} {Disk} and {Constraints} on the {Presence} of {Giant} {Planets} in the {Habitable} {Zone} of β {Leo}},
	volume = {161},
	issn = {1538-3881},
	shorttitle = {The {HOSTS} {Survey}},
	url = {https://doi.org/10.3847/1538-3881/abe3ff},
	doi = {10.3847/1538-3881/abe3ff},
	language = {en},
	number = {4},
	urldate = {2026-08-03},
	journal = {The Astronomical Journal},
	publisher = {The American Astronomical Society},
	author = {Defrère, D. and Hinz, P. M. and Kennedy, G. M. and Stone, J. and Rigley, J. and Ertel, S. and Gaspar, A. and Bailey, V. P. and Hoffmann, W. F. and Mennesson, B. and Millan-Gabet, R. and Danchi, W. C. and Absil, O. and Arbo, P. and Beichman, C and Bonavita, M and Brusa, G. and Bryden, G. and Downey, E. C. and Esposito, S. and Grenz, P. and Haniff, C. and Hill, J. M. and Leisenring, J. M. and Males, J. R. and McMahon, T. J. and Montoya, M. and Morzinski, K. M. and Pinna, E. and Puglisi, A. and Rieke, G. and Roberge, A. and Rousseau, H. and Serabyn, E. and Spalding, E. and Skemer, A. J. and Stapelfeldt, K. and Su, K. and Vaz, A. and Weinberger, A. J. and Wyatt, M. C.},
	month = mar,
	year = {2021},
	pages = {186},
}

@article{stark_maximizing_2014,
	title = {{MAXIMIZING} {THE} {ExoEarth} {CANDIDATE} {YIELD} {FROM} {A} {FUTURE} {DIRECT} {IMAGING} {MISSION}},
	volume = {795},
	issn = {0004-637X},
	url = {https://doi.org/10.1088/0004-637X/795/2/122},
	doi = {10.1088/0004-637X/795/2/122},
	language = {en},
	number = {2},
	urldate = {2026-08-03},
	journal = {The Astrophysical Journal},
	publisher = {The American Astronomical Society},
	author = {Stark, Christopher C. and Roberge, Aki and Mandell, Avi and Robinson, Tyler D.},
	month = oct,
	year = {2014},
	pages = {122},
}

@article{roberge_exozodiacal_2012,
	title = {The {Exozodiacal} {Dust} {Problem} for {Direct} {Observations} of {Exo}-{Earths}},
	volume = {124},
	issn = {1538-3873},
	url = {https://iopscience.iop.org/article/10.1086/667218},
	doi = {10.1086/667218},
	language = {en},
	number = {918},
	urldate = {2026-08-03},
	journal = {Publications of the Astronomical Society of the Pacific},
	publisher = {IOP Publishing},
	author = {Roberge, Aki and Chen, Christine H. and Millan-Gabet, Rafael and Weinberger, Alycia J. and Hinz, Philip M. and Stapelfeldt, Karl R. and Absil, Olivier and Kuchner, Marc J. and Bryden, Geoffrey},
	month = aug,
	year = {2012},
	pages = {799},
}

@article{defrere_nulling_2010,
	title = {Nulling interferometry: impact of exozodiacal clouds on the performance of future life-finding space missions},
	volume = {509},
	copyright = {© ESO, 2010},
	issn = {0004-6361, 1432-0746},
	shorttitle = {Nulling interferometry},
	url = {https://www.aanda.org/articles/aa/abs/2010/01/aa12973-09/aa12973-09.html},
	doi = {10.1051/0004-6361/200912973},
	language = {en},
	urldate = {2026-08-03},
	journal = {Astronomy \& Astrophysics},
	publisher = {EDP Sciences},
	author = {Defrère, D. and Absil, O. and Hartog, R. Den and Hanot, C. and Stark, C.},
	month = jan,
	year = {2010},
	pages = {A9},
}

@article{defrere_first-light_2015,
	title = {{FIRST}-{LIGHT} {LBT} {NULLING} {INTERFEROMETRIC} {OBSERVATIONS}: {WARM} {EXOZODIACAL} {DUST} {RESOLVED} {WITHIN} {A} {FEW} {AU} {OF} η {Crv}},
	volume = {799},
	issn = {0004-637X},
	shorttitle = {{FIRST}-{LIGHT} {LBT} {NULLING} {INTERFEROMETRIC} {OBSERVATIONS}},
	url = {https://doi.org/10.1088/0004-637X/799/1/42},
	doi = {10.1088/0004-637X/799/1/42},
	language = {en},
	number = {1},
	urldate = {2026-08-03},
	journal = {The Astrophysical Journal},
	publisher = {The American Astronomical Society},
	author = {Defrère, D. and Hinz, P. M. and Skemer, A. J. and Kennedy, G. M. and Bailey, V. P. and Hoffmann, W. F. and Mennesson, B. and Millan-Gabet, R. and Danchi, W. C. and Absil, O. and Arbo, P. and Beichman, C. and Brusa, G. and Bryden, G. and Downey, E. C. and Durney, O. and Esposito, S. and Gaspar, A. and Grenz, P. and Haniff, C. and Hill, J. M. and Lebreton, J. and Leisenring, J. M. and Males, J. R. and Marion, L. and McMahon, T. J. and Montoya, M. and Morzinski, K. M. and Pinna, E. and Puglisi, A. and Rieke, G. and Roberge, A. and Serabyn, E. and Sosa, R. and Stapeldfeldt, K. and Su, K. and Vaitheeswaran, V. and Vaz, A. and Weinberger, A. J. and Wyatt, M. C.},
	month = jan,
	year = {2015},
	pages = {42},
}

@article{mennesson_constraining_2014,
	title = {{CONSTRAINING} {THE} {EXOZODIACAL} {LUMINOSITY} {FUNCTION} {OF} {MAIN}-{SEQUENCE} {STARS}: {COMPLETE} {RESULTS} {FROM} {THE} {KECK} {NULLER} {MID}-{INFRARED} {SURVEYS}},
	volume = {797},
	issn = {0004-637X},
	shorttitle = {{CONSTRAINING} {THE} {EXOZODIACAL} {LUMINOSITY} {FUNCTION} {OF} {MAIN}-{SEQUENCE} {STARS}},
	url = {https://doi.org/10.1088/0004-637X/797/2/119},
	doi = {10.1088/0004-637X/797/2/119},
	language = {en},
	number = {2},
	urldate = {2026-08-03},
	journal = {The Astrophysical Journal},
	publisher = {The American Astronomical Society},
	author = {Mennesson, B. and Millan-Gabet, R. and Serabyn, E. and Colavita, M. M. and Absil, O. and Bryden, G. and Wyatt, M. and Danchi, W. and Defrère, D. and Doré, O. and Hinz, P. and Kuchner, M. and Ragland, S. and Scott, N. and Stapelfeldt, K. and Traub, W. and Woillez, J.},
	month = dec,
	year = {2014},
	pages = {119},
}

@article{rigley_comet_2022,
	title = {Comet fragmentation as a source of the zodiacal cloud},
	volume = {510},
	issn = {0035-8711},
	url = {https://doi.org/10.1093/mnras/stab3482},
	doi = {10.1093/mnras/stab3482},
	number = {1},
	urldate = {2026-08-03},
	journal = {Monthly Notices of the Royal Astronomical Society},
	author = {Rigley, Jessica K and Wyatt, Mark C},
	month = feb,
	year = {2022},
	pages = {834--857},
}

@article{faramaz_inner_2017,
	title = {Inner mean-motion resonances with eccentric planets: a possible origin for exozodiacal dust clouds},
	volume = {465},
	issn = {0035-8711},
	shorttitle = {Inner mean-motion resonances with eccentric planets},
	url = {https://ui.adsabs.harvard.edu/abs/2017MNRAS.465.2352F},
	doi = {10.1093/mnras/stw2846},
	urldate = {2026-08-03},
	journal = {Monthly Notices of the Royal Astronomical Society},
	publisher = {OUP},
	author = {Faramaz, V. and Ertel, S. and Booth, M. and Cuadra, J. and Simmonds, C.},
	month = feb,
	year = {2017},
	note = {ADS Bibcode: 2017MNRAS.465.2352F},
	pages = {2352--2365},
}

@article{marboeuf_extrasolar_2016,
	series = {Cosmic {Dust} {VIII}},
	title = {Extrasolar comets: {The} origin of dust in exozodiacal disks?},
	volume = {133},
	issn = {0032-0633},
	shorttitle = {Extrasolar comets},
	url = {https://www.sciencedirect.com/science/article/pii/S003206331530101X},
	doi = {10.1016/j.pss.2016.03.014},
	urldate = {2026-08-03},
	journal = {Planetary and Space Science},
	author = {Marboeuf, U. and Bonsor, A. and Augereau, J. -C.},
	month = nov,
	year = {2016},
	pages = {47--62},
}

@article{kennedy_warm_2015,
	title = {Warm exo-{Zodi} from cool exo-{Kuiper} belts: the significance of {P}–{R} drag and the inference of intervening planets},
	volume = {449},
	issn = {0035-8711},
	shorttitle = {Warm exo-{Zodi} from cool exo-{Kuiper} belts},
	url = {https://doi.org/10.1093/mnras/stv453},
	doi = {10.1093/mnras/stv453},
	number = {3},
	urldate = {2026-08-03},
	journal = {Monthly Notices of the Royal Astronomical Society},
	author = {Kennedy, Grant M. and Piette, Anjali},
	month = may,
	year = {2015},
	pages = {2304--2311},
}

@article{garreau_hosts_2025,
	title = {The {HOSTS} survey: {Suspected} variable dust emission and constraints on companions around θ {Boo}},
	volume = {699},
	copyright = {© The Authors 2025},
	issn = {0004-6361, 1432-0746},
	shorttitle = {The {HOSTS} survey},
	url = {https://www.aanda.org/articles/aa/abs/2025/07/aa52653-24/aa52653-24.html},
	doi = {10.1051/0004-6361/202452653},
	language = {en},
	urldate = {2026-07-28},
	journal = {Astronomy \& Astrophysics},
	publisher = {EDP Sciences},
	author = {Garreau, G. and Defrère, D. and Ertel, S. and Faramaz-Gorka, V. and Bryden, G. and Sommer, M. and Mesa, D. and Wagner, K. and Prins, T. De and Laugier, R. and Weinberger, A. and Farinato, J. and Haniff, C. and Hinz, P. M. and Isbell, J. W. and Kennedy, G. M. and Lorenzetto, A. and Maier, E. R. and Marafatto, L. and Marino, S. and Martinod, M. A. and Mennesson, B. and Rousseau, H. and Spalding, E. and Vassallo, D. and Wyatt, M. C.},
	month = jul,
	year = {2025},
	pages = {A107},
}

@article{ohno_grain_2021,
	title = {Grain {Growth} in {Escaping} {Atmospheres}: {Implications} for the {Radius} {Inflation} of {Super}-{Puffs}},
	volume = {920},
	issn = {0004-637X},
	shorttitle = {Grain {Growth} in {Escaping} {Atmospheres}},
	url = {https://doi.org/10.3847/1538-4357/ac1516},
	doi = {10.3847/1538-4357/ac1516},
	language = {en},
	number = {2},
	urldate = {2026-07-19},
	journal = {The Astrophysical Journal},
	publisher = {The American Astronomical Society},
	author = {Ohno, Kazumasa and Tanaka, Yuki A.},
	month = oct,
	year = {2021},
	pages = {124},
}

@article{plane_atmospheric_2003,
	title = {Atmospheric {Chemistry} of {Meteoric} {Metals}},
	volume = {103},
	issn = {0009-2665},
	url = {https://doi.org/10.1021/cr0205309},
	doi = {10.1021/cr0205309},
	number = {12},
	urldate = {2026-07-19},
	journal = {Chemical Reviews},
	publisher = {American Chemical Society},
	author = {Plane, John M. C.},
	month = dec,
	year = {2003},
	pages = {4963--4984},
}

@article{daly_meteoric_2020,
	title = {The {Meteoric} {Ni} {Layer} in the {Upper} {Atmosphere}},
	volume = {125},
	copyright = {©2020. The Authors.},
	issn = {2169-9402},
	url = {https://onlinelibrary.wiley.com/doi/abs/10.1029/2020JA028083},
	doi = {10.1029/2020JA028083},
	language = {en},
	number = {8},
	urldate = {2026-07-19},
	journal = {Journal of Geophysical Research: Space Physics},
	author = {Daly, Shane M. and Feng, Wuhu and Mangan, Thomas P. and Gerding, Michael and Plane, John M. C.},
	year = {2020},
	note = {\_eprint: https://agupubs.onlinelibrary.wiley.com/doi/pdf/10.1029/2020JA028083},
	pages = {e2020JA028083},
}

@misc{attaway_mienet_2026,
	title = {{MieNet}: {Fast} {Opacity} {Calculations} of {Well}-mixed {Cloud} {Particles}},
	shorttitle = {{MieNet}},
	url = {https://zenodo.org/records/20516973},
	doi = {10.5281/zenodo.20516973},
	urldate = {2026-06-02},
	publisher = {Zenodo},
	author = {Attaway, Daisy and Kiefer, Sven and Zhao, Yinan and Morley, Caroline},
	month = jun,
	year = {2026},
}

@article{gulick_end_2025,
	title = {End of the {Cretaceous}},
	volume = {544},
	url = {https://www.lyellcollection.org/doi/full/10.1144/SP544-2023-176},
	doi = {10.1144/SP544-2023-176},
	number = {1},
	urldate = {2026-05-29},
	journal = {Geological Society, London, Special Publications},
	publisher = {The Geological Society of London},
	author = {Gulick, Sean P. S.},
	month = mar,
	year = {2025},
	pages = {549--570},
}

@inproceedings{kulik_results_1940,
	title = {Results on the {Tunguska} meteorite, collected up until 1939},
	volume = {22},
	booktitle = {Reports of the {Academy} of {Science} of the {U}. {S}. {S}. {R}.},
	author = {Kulik, L. A.},
	year = {1940},
	pages = {520--524},
}

@article{popova_chelyabinsk_2013,
	title = {Chelyabinsk {Airburst}, {Damage} {Assessment}, {Meteorite} {Recovery}, and {Characterization}},
	volume = {342},
	url = {https://www.science.org/doi/full/10.1126/science.1242642},
	doi = {10.1126/science.1242642},
	number = {6162},
	urldate = {2026-05-26},
	journal = {Science},
	publisher = {American Association for the Advancement of Science},
	author = {Popova, Olga P. and Jenniskens, Peter and Emel’yanenko, Vacheslav and Kartashova, Anna and Biryukov, Eugeny and Khaibrakhmanov, Sergey and Shuvalov, Valery and Rybnov, Yurij and Dudorov, Alexandr and Grokhovsky, Victor I. and Badyukov, Dmitry D. and Yin, Qing-Zhu and Gural, Peter S. and Albers, Jim and Granvik, Mikael and Evers, Läslo G. and Kuiper, Jacob and Kharlamov, Vladimir and Solovyov, Andrey and Rusakov, Yuri S. and Korotkiy, Stanislav and Serdyuk, Ilya and Korochantsev, Alexander V. and Larionov, Michail Yu. and Glazachev, Dmitry and Mayer, Alexander E. and Gisler, Galen and Gladkovsky, Sergei V. and Wimpenny, Josh and Sanborn, Matthew E. and Yamakawa, Akane and Verosub, Kenneth L. and Rowland, Douglas J. and Roeske, Sarah and Botto, Nicholas W. and Friedrich, Jon M. and Zolensky, Michael E. and Le, Loan and Ross, Daniel and Ziegler, Karen and Nakamura, Tomoki and Ahn, Insu and Lee, Jong Ik and Zhou, Qin and Li, Xian-Hua and Li, Qiu-Li and Liu, Yu and Tang, Guo-Qiang and Hiroi, Takahiro and Sears, Derek and Weinstein, Ilya A. and Vokhmintsev, Alexander S. and Ishchenko, Alexei V. and Schmitt-Kopplin, Phillipe and Hertkorn, Norbert and Nagao, Keisuke and Haba, Makiko K. and Komatsu, Mutsumi and Mikouchi, Takashi and {(THE CHELYABINSK AIRBURST CONSORTIUM)}},
	month = nov,
	year = {2013},
	pages = {1069--1073},
}

@article{kyte_accretion_1986,
	title = {Accretion {Rate} of {Extraterrestrial} {Matter}: {Iridium} {Deposited} 33 to 67 {Million} {Years} {Ago}},
	volume = {232},
	shorttitle = {Accretion {Rate} of {Extraterrestrial} {Matter}},
	url = {https://www.science.org/doi/10.1126/science.232.4755.1225},
	doi = {10.1126/science.232.4755.1225},
	number = {4755},
	urldate = {2026-05-26},
	journal = {Science},
	publisher = {American Association for the Advancement of Science},
	author = {Kyte, Frank T. and Wasson, John T.},
	month = jun,
	year = {1986},
	pages = {1225--1229},
}

@article{ertel_hosts_2020,
	title = {The {HOSTS} {Survey} for {Exozodiacal} {Dust}: {Observational} {Results} from the {Complete} {Survey}},
	volume = {159},
	issn = {1538-3881},
	shorttitle = {The {HOSTS} {Survey} for {Exozodiacal} {Dust}},
	url = {https://doi.org/10.3847/1538-3881/ab7817},
	doi = {10.3847/1538-3881/ab7817},
	language = {en},
	number = {4},
	urldate = {2026-05-21},
	journal = {The Astronomical Journal},
	publisher = {The American Astronomical Society},
	author = {Ertel, S. and Defrère, D. and Hinz, P. and Mennesson, B. and Kennedy, G. M. and Danchi, W. C. and Gelino, C. and Hill, J. M. and Hoffmann, W. F. and Mazoyer, J. and Rieke, G. and Shannon, A. and Stapelfeldt, K. and Spalding, E. and Stone, J. M. and Vaz, A. and Weinberger, A. J. and Willems, P. and Absil, O. and Arbo, P. and Bailey, V. P. and Beichman, C. and Bryden, G. and Downey, E. C. and Durney, O. and Esposito, S. and Gaspar, A. and Grenz, P. and Haniff, C. A. and Leisenring, J. M. and Marion, L. and McMahon, T. J. and Millan-Gabet, R. and Montoya, M. and Morzinski, K. M. and Perera, S. and Pinna, E. and Pott, J.-U. and Power, J. and Puglisi, A. and Roberge, A. and Serabyn, E. and Skemer, A. J. and Su, K. Y. L. and Vaitheeswaran, V. and Wyatt, M. C.},
	month = mar,
	year = {2020},
	pages = {177},
}

@article{ohno_super-rayleigh_2020,
	title = {Super-{Rayleigh} {Slopes} in {Transmission} {Spectra} of {Exoplanets} {Generated} by {Photochemical} {Haze}},
	volume = {895},
	issn = {0004-637X},
	url = {https://ui.adsabs.harvard.edu/abs/2020ApJ...895L..47O},
	doi = {10.3847/2041-8213/ab93d7},
	urldate = {2026-05-15},
	journal = {The Astrophysical Journal},
	publisher = {IOP},
	author = {Ohno, Kazumasa and Kawashima, Yui},
	month = jun,
	year = {2020},
	note = {ADS Bibcode: 2020ApJ...895L..47O},
	pages = {L47},
}

@article{lodge_fractal_2026,
	title = {Fractal {Aggregate} {Aerosols} in the {Virga} {Cloud} {Code}. {II}. {Exploring} the {Effects} of {Key} {Cloud} {Parameters} in {Warm} {Neptune}, {Hot} {Jupiter} and {Brown} {Dwarf} {Atmospheres}},
	volume = {997},
	issn = {0004-637X},
	url = {https://ui.adsabs.harvard.edu/abs/2026ApJ...997..317L},
	doi = {10.3847/1538-4357/ae2752},
	urldate = {2026-05-09},
	journal = {The Astrophysical Journal},
	publisher = {IOP},
	author = {Lodge, Matt G. and Moran, Sarah E. and Wakeford, Hannah R. and Leinhardt, Zoë M. and Marley, Mark S.},
	month = feb,
	year = {2026},
	note = {ADS Bibcode: 2026ApJ...997..317L},
	pages = {317},
}

@article{frantseva_delivery_2018,
	title = {Delivery of organics to {Mars} through asteroid and comet impacts},
	volume = {309},
	issn = {0019-1035},
	url = {https://www.sciencedirect.com/science/article/pii/S0019103517304335},
	doi = {10.1016/j.icarus.2018.03.006},
	urldate = {2026-03-14},
	journal = {Icarus},
	author = {Frantseva, Kateryna and Mueller, Michael and ten Kate, Inge Loes and van der Tak, Floris F. S. and Greenstreet, Sarah},
	month = jul,
	year = {2018},
	pages = {125--133},
}

@article{cziczo_ablation_2001,
	title = {Ablation, {Flux}, and {Atmospheric} {Implications} of {Meteors} {Inferred} from {Stratospheric} {Aerosol}},
	volume = {291},
	url = {https://www.science.org/doi/full/10.1126/science.1057737},
	doi = {10.1126/science.1057737},
	number = {5509},
	urldate = {2025-12-17},
	journal = {Science},
	publisher = {American Association for the Advancement of Science},
	author = {Cziczo, D. J. and Thomson, D. S. and Murphy, D. M.},
	month = mar,
	year = {2001},
	pages = {1772--1775},
}

@article{bell_nightside_2024,
	title = {Nightside clouds and disequilibrium chemistry on the hot {Jupiter} {WASP}-43b},
	volume = {8},
	copyright = {2024 The Author(s)},
	issn = {2397-3366},
	url = {https://www.nature.com/articles/s41550-024-02230-x},
	doi = {10.1038/s41550-024-02230-x},
	language = {en},
	number = {7},
	urldate = {2026-02-23},
	journal = {Nature Astronomy},
	publisher = {Nature Publishing Group},
	author = {Bell, Taylor J. and Crouzet, Nicolas and Cubillos, Patricio E. and Kreidberg, Laura and Piette, Anjali A. A. and Roman, Michael T. and Barstow, Joanna K. and Blecic, Jasmina and Carone, Ludmila and Coulombe, Louis-Philippe and Ducrot, Elsa and Hammond, Mark and Mendonça, João M. and Moses, Julianne I. and Parmentier, Vivien and Stevenson, Kevin B. and Teinturier, Lucas and Zhang, Michael and Batalha, Natalie M. and Bean, Jacob L. and Benneke, Björn and Charnay, Benjamin and Chubb, Katy L. and Demory, Brice-Olivier and Gao, Peter and Lee, Elspeth K. H. and López-Morales, Mercedes and Morello, Giuseppe and Rauscher, Emily and Sing, David K. and Tan, Xianyu and Venot, Olivia and Wakeford, Hannah R. and Aggarwal, Keshav and Ahrer, Eva-Maria and Alam, Munazza K. and Baeyens, Robin and Barrado, David and Caceres, Claudio and Carter, Aarynn L. and Casewell, Sarah L. and Challener, Ryan C. and Crossfield, Ian J. M. and Decin, Leen and Désert, Jean-Michel and Dobbs-Dixon, Ian and Dyrek, Achrène and Espinoza, Néstor and Feinstein, Adina D. and Gibson, Neale P. and Harrington, Joseph and Helling, Christiane and Hu, Renyu and Iro, Nicolas and Kempton, Eliza M.-R. and Kendrew, Sarah and Komacek, Thaddeus D. and Krick, Jessica and Lagage, Pierre-Olivier and Leconte, Jérémy and Lendl, Monika and Lewis, Neil T. and Lothringer, Joshua D. and Malsky, Isaac and Mancini, Luigi and Mansfield, Megan and Mayne, Nathan J. and Evans-Soma, Thomas M. and Molaverdikhani, Karan and Nikolov, Nikolay K. and Nixon, Matthew C. and Palle, Enric and Petit dit de la Roche, Dominique J. M. and Piaulet, Caroline and Powell, Diana and Rackham, Benjamin V. and Schneider, Aaron D. and Steinrueck, Maria E. and Taylor, Jake and Welbanks, Luis and Yurchenko, Sergei N. and Zhang, Xi and Zieba, Sebastian},
	month = jul,
	year = {2024},
	pages = {879--898},
}

@article{bardeen_numerical_2008,
	title = {Numerical simulations of the three-dimensional distribution of meteoric dust in the mesosphere and upper stratosphere},
	volume = {113},
	copyright = {Copyright 2008 by the American Geophysical Union.},
	issn = {2156-2202},
	url = {https://onlinelibrary.wiley.com/doi/abs/10.1029/2007JD009515},
	doi = {10.1029/2007JD009515},
	language = {en},
	number = {D17},
	urldate = {2023-12-01},
	journal = {J. Geophys. Res. Atmos.},
	author = {Bardeen, C. G. and Toon, O. B. and Jensen, E. J. and Marsh, D. R. and Harvey, V. L.},
	year = {2008},
}

@article{min_scattering_2003,
	series = {Electromagnetic and {Light} {Scattering} by {Non}-{Spherical} {Particles}},
	title = {Scattering and absorption cross sections for randomly oriented spheroids of arbitrary size},
	volume = {79-80},
	issn = {0022-4073},
	url = {https://www.sciencedirect.com/science/article/pii/S0022407302003308},
	doi = {10.1016/S0022-4073(02)00330-8},
	urldate = {2026-05-03},
	journal = {Journal of Quantitative Spectroscopy and Radiative Transfer},
	author = {Min, M and Hovenier, J. W and de Koter, A},
	month = jun,
	year = {2003},
	pages = {939--951},
}

@article{rojas_micrometeorite_2021,
	title = {The micrometeorite flux at {Dome} {C} ({Antarctica}), monitoring the accretion of extraterrestrial dust on {Earth}},
	volume = {560},
	issn = {0012-821X},
	url = {https://www.sciencedirect.com/science/article/pii/S0012821X21000534},
	doi = {10.1016/j.epsl.2021.116794},
	urldate = {2026-04-28},
	journal = {Earth and Planetary Science Letters},
	author = {Rojas, J. and Duprat, J. and Engrand, C. and Dartois, E. and Delauche, L. and Godard, M. and Gounelle, M. and Carrillo-Sánchez, J. D. and Pokorný, P. and Plane, J. M. C.},
	month = apr,
	year = {2021},
	pages = {116794},
}

@article{drolshagen_mass_2017,
	series = {{SI}:{Meteoroids} 2016},
	title = {Mass accumulation of earth from interplanetary dust, meteoroids, asteroids and comets},
	volume = {143},
	issn = {0032-0633},
	url = {https://www.sciencedirect.com/science/article/pii/S0032063316302434},
	doi = {10.1016/j.pss.2016.12.010},
	urldate = {2026-04-28},
	journal = {Planetary and Space Science},
	author = {Drolshagen, Gerhard and Koschny, Detlef and Drolshagen, Sandra and Kretschmer, Jana and Poppe, Björn},
	month = sep,
	year = {2017},
	pages = {21--27},
}

@article{wasson_comment_1987,
	title = {Comment on the letter “{On} the influx of small comets into the {Earth}'s atmosphere {II}: {Interpretation}”},
	volume = {14},
	copyright = {Copyright 1986 by the American Geophysical Union.},
	issn = {1944-8007},
	shorttitle = {Comment on the letter “{On} the influx of small comets into the {Earth}'s atmosphere {II}},
	url = {https://onlinelibrary.wiley.com/doi/abs/10.1029/GL014i007p00779},
	doi = {10.1029/GL014i007p00779},
	language = {en},
	number = {7},
	urldate = {2026-04-28},
	journal = {Geophysical Research Letters},
	author = {Wasson, John T. and Kyte, Frank T.},
	year = {1987},
	note = {\_eprint: https://agupubs.onlinelibrary.wiley.com/doi/pdf/10.1029/GL014i007p00779},
	pages = {779--780},
}

@article{peucker-ehrenbrink_accretion_1996,
	title = {Accretion of extraterrestrial matter during the last 80 million years and its effect on the marine osmium isotope record},
	volume = {60},
	issn = {0016-7037},
	url = {https://www.sciencedirect.com/science/article/pii/0016703796001615},
	doi = {10.1016/0016-7037(96)00161-5},
	number = {17},
	urldate = {2026-04-28},
	journal = {Geochimica et Cosmochimica Acta},
	author = {Peucker-Ehrenbrink, B.},
	month = sep,
	year = {1996},
	pages = {3187--3196},
}

@article{gabrielli_meteoric_2004,
	title = {Meteoric smoke fallout over the {Holocene} epoch revealed by iridium and platinum in {Greenland} ice},
	volume = {432},
	copyright = {2005 Macmillan Magazines Ltd.},
	issn = {1476-4687},
	url = {https://www.nature.com/articles/nature03137},
	doi = {10.1038/nature03137},
	language = {en},
	number = {7020},
	urldate = {2026-04-28},
	journal = {Nature},
	publisher = {Nature Publishing Group},
	author = {Gabrielli, Paolo and Barbante, Carlo and Plane, John M. C. and Varga, Anita and Hong, Sungmin and Cozzi, Giulio and Gaspari, Vania and Planchon, Frédéric A. M. and Cairns, Warren and Ferrari, Christophe and Crutzen, Paul and Cescon, Paolo and Boutron, Claude F.},
	month = dec,
	year = {2004},
	pages = {1011--1014},
}

@article{lanci_meteoric_2006,
	title = {Meteoric smoke fallout revealed by superparamagnetism in {Greenland} ice},
	volume = {33},
	issn = {1944-8007},
	url = {https://agupubs.onlinelibrary.wiley.com/doi/10.1029/2006GL026480},
	doi = {10.1029/2006GL026480},
	language = {en},
	number = {13},
	urldate = {2026-04-28},
	journal = {Geophysical Research Letters},
	publisher = {John Wiley \& Sons, Ltd},
	author = {Lanci, L. and Kent, D. V.},
	month = jul,
	year = {2006},
}

@article{lanci_meteoric_2007,
	title = {Meteoric smoke concentration in the {Vostok} ice core estimated from superparamagnetic relaxation and some consequences for estimates of {Earth} accretion rate},
	volume = {34},
	copyright = {Copyright 2007 by the American Geophysical Union.},
	issn = {1944-8007},
	url = {https://onlinelibrary.wiley.com/doi/abs/10.1029/2007GL029811},
	doi = {10.1029/2007GL029811},
	language = {en},
	number = {10},
	urldate = {2026-04-28},
	journal = {Geophysical Research Letters},
	author = {Lanci, Luca and Kent, Dennis V. and Biscaye, Pierre E.},
	year = {2007},
	note = {\_eprint: https://agupubs.onlinelibrary.wiley.com/doi/pdf/10.1029/2007GL029811},
}

@article{hervig_first_2009,
	title = {First {Satellite} {Observations} of {Meteoric} {Smoke} in the {Middle} {Atmosphere}},
	volume = {36},
	copyright = {Copyright 2009 by the American Geophysical Union.},
	issn = {1944-8007},
	url = {https://onlinelibrary.wiley.com/doi/abs/10.1029/2009GL039737},
	doi = {10.1029/2009GL039737},
	language = {en},
	number = {18},
	urldate = {2026-04-28},
	journal = {Geophysical Research Letters},
	author = {Hervig, Mark E. and Gordley, Larry L. and Deaver, Lance E. and Siskind, David E. and Stevens, Michael H. and Russell III, James M. and Bailey, Scott M. and Megner, Linda and Bardeen, Charles G.},
	year = {2009},
	note = {\_eprint: https://agupubs.onlinelibrary.wiley.com/doi/pdf/10.1029/2009GL039737},
}

@article{gardner_seasonal_2011,
	title = {Seasonal variations of the mesospheric {Fe} layer at {Rothera}, {Antarctica} (67.5°{S}, 68.0°{W})},
	volume = {116},
	copyright = {Copyright 2011 by the American Geophysical Union.},
	issn = {2156-2202},
	url = {https://onlinelibrary.wiley.com/doi/abs/10.1029/2010JD014655},
	doi = {10.1029/2010JD014655},
	language = {en},
	number = {D2},
	urldate = {2026-04-28},
	journal = {Journal of Geophysical Research: Atmospheres},
	author = {Gardner, Chester S. and Chu, Xinzhao and Espy, Patrick J. and Plane, John M. C. and Marsh, Daniel R. and Janches, Diego},
	year = {2011},
	note = {\_eprint: https://agupubs.onlinelibrary.wiley.com/doi/pdf/10.1029/2010JD014655},
}

@article{plane_time-resolved_2004,
	title = {A time-resolved model of the mesospheric {Na} layer: constraints on the meteor input function},
	volume = {4},
	issn = {1680-7316},
	shorttitle = {A time-resolved model of the mesospheric {Na} layer},
	url = {https://acp.copernicus.org/articles/4/627/2004/},
	doi = {10.5194/acp-4-627-2004},
	language = {English},
	number = {3},
	urldate = {2026-04-28},
	journal = {Atmospheric Chemistry and Physics},
	publisher = {Copernicus GmbH},
	author = {Plane, J. M. C.},
	month = apr,
	year = {2004},
	pages = {627--638},
}

@article{mathews_micrometeoroid_2001,
	title = {The micrometeoroid mass flux into the upper atmosphere: {Arecibo} results and a comparison with prior estimates},
	volume = {28},
	copyright = {Copyright 2001 by the American Geophysical Union.},
	issn = {1944-8007},
	shorttitle = {The micrometeoroid mass flux into the upper atmosphere},
	url = {https://onlinelibrary.wiley.com/doi/abs/10.1029/2000GL012621},
	doi = {10.1029/2000GL012621},
	language = {en},
	number = {10},
	urldate = {2026-04-28},
	journal = {Geophysical Research Letters},
	author = {Mathews, J. D. and Janches, D. and Meisel, D. D. and Zhou, Q -H.},
	year = {2001},
	note = {\_eprint: https://agupubs.onlinelibrary.wiley.com/doi/pdf/10.1029/2000GL012621},
	pages = {1929--1932},
}

@article{nesvorny_cometary_2010,
	title = {{COMETARY} {ORIGIN} {OF} {THE} {ZODIACAL} {CLOUD} {AND} {CARBONACEOUS} {MICROMETEORITES}. {IMPLICATIONS} {FOR} {HOT} {DEBRIS} {DISKS}},
	volume = {713},
	issn = {0004-637X},
	url = {https://doi.org/10.1088/0004-637X/713/2/816},
	doi = {10.1088/0004-637X/713/2/816},
	language = {en},
	number = {2},
	urldate = {2026-04-28},
	journal = {The Astrophysical Journal},
	publisher = {The American Astronomical Society},
	author = {Nesvorný, David and Jenniskens, Peter and Levison, Harold F. and Bottke, William F. and Vokrouhlický, David and Gounelle, Matthieu},
	month = mar,
	year = {2010},
	pages = {816},
}

@article{gaspar_collisional_2013,
	title = {{THE} {COLLISIONAL} {EVOLUTION} {OF} {DEBRIS} {DISKS}},
	volume = {768},
	issn = {0004-637X},
	url = {https://doi.org/10.1088/0004-637X/768/1/25},
	doi = {10.1088/0004-637X/768/1/25},
	language = {en},
	number = {1},
	urldate = {2026-04-21},
	journal = {The Astrophysical Journal},
	publisher = {The American Astronomical Society},
	author = {Gáspár, András and Rieke, George H. and Balog, Zoltán},
	month = apr,
	year = {2013},
	pages = {25},
}

@article{kreidberg_first_2025,
	title = {A first look at rocky exoplanets with {JWST}},
	volume = {122},
	url = {https://www.pnas.org/doi/abs/10.1073/pnas.2416190122},
	doi = {10.1073/pnas.2416190122},
	number = {39},
	urldate = {2026-04-17},
	journal = {Proceedings of the National Academy of Sciences},
	publisher = {Proceedings of the National Academy of Sciences},
	author = {Kreidberg, Laura and Stevenson, Kevin B.},
	month = sep,
	year = {2025},
	pages = {e2416190122},
}

@article{moran_high_2023,
	title = {High {Tide} or {Riptide} on the {Cosmic} {Shoreline}? {A} {Water}-rich {Atmosphere} or {Stellar} {Contamination} for the {Warm} {Super}-{Earth} {GJ} 486b from {JWST} {Observations}},
	volume = {948},
	issn = {2041-8205},
	shorttitle = {High {Tide} or {Riptide} on the {Cosmic} {Shoreline}?},
	url = {https://doi.org/10.3847/2041-8213/accb9c},
	doi = {10.3847/2041-8213/accb9c},
	language = {en},
	number = {1},
	urldate = {2026-04-17},
	journal = {The Astrophysical Journal Letters},
	publisher = {The American Astronomical Society},
	author = {Moran, Sarah E. and Stevenson, Kevin B. and Sing, David K. and MacDonald, Ryan J. and Kirk, James and Lustig-Yaeger, Jacob and Peacock, Sarah and Mayorga, L. C. and Bennett, Katherine A. and López-Morales, Mercedes and May, E. M. and Rustamkulov, Zafar and Valenti, Jeff A. and Adams Redai, Jéa I. and Alam, Munazza K. and Batalha, Natasha E. and Fu, Guangwei and Gonzalez-Quiles, Junellie and Highland, Alicia N. and Kruse, Ethan and Lothringer, Joshua D. and Ortiz Ceballos, Kevin N. and Sotzen, Kristin S. and Wakeford, Hannah R.},
	month = may,
	year = {2023},
	pages = {L11},
}

@article{may_double_2023,
	title = {Double {Trouble}: {Two} {Transits} of the {Super}-{Earth} {GJ} 1132 b {Observed} with {JWST} {NIRSpec} {G395H}},
	volume = {959},
	issn = {2041-8205},
	shorttitle = {Double {Trouble}},
	url = {https://doi.org/10.3847/2041-8213/ad054f},
	doi = {10.3847/2041-8213/ad054f},
	language = {en},
	number = {1},
	urldate = {2026-04-17},
	journal = {The Astrophysical Journal Letters},
	publisher = {The American Astronomical Society},
	author = {May, E. M. and MacDonald, Ryan J. and Bennett, Katherine A. and Moran, Sarah E. and Wakeford, Hannah R. and Peacock, Sarah and Lustig-Yaeger, Jacob and Highland, Alicia N. and Stevenson, Kevin B. and Sing, David K. and Mayorga, L. C. and Batalha, Natasha E. and Kirk, James and López-Morales, Mercedes and Valenti, Jeff A. and Alam, Munazza K. and Alderson, Lili and Fu, Guangwei and Gonzalez-Quiles, Junellie and Lothringer, Joshua D. and Rustamkulov, Zafar and Sotzen, Kristin S.},
	month = dec,
	year = {2023},
	pages = {L9},
}

@article{scarsdale_jwst_2024,
	title = {{JWST} {COMPASS}: {The} 3–5 μm {Transmission} {Spectrum} of the {Super}-{Earth} {L} 98-59 c},
	volume = {168},
	issn = {1538-3881},
	shorttitle = {{JWST} {COMPASS}},
	url = {https://doi.org/10.3847/1538-3881/ad73cf},
	doi = {10.3847/1538-3881/ad73cf},
	language = {en},
	number = {6},
	urldate = {2026-04-17},
	journal = {The Astronomical Journal},
	publisher = {The American Astronomical Society},
	author = {Scarsdale, Nicholas and Wogan, Nicholas and Wakeford, Hannah R. and Wallack, Nicole L. and Batalha, Natasha E. and Alderson, Lili and Aguichine, Artyom and Wolfgang, Angie and Teske, Johanna and Moran, Sarah E. and López-Morales, Mercedes and Kirk, James and Gordon, Tyler and Gao, Peter and Batalha, Natalie M. and Alam, Munazza K. and Adams Redai, Jea},
	month = nov,
	year = {2024},
	pages = {276},
}

@article{kirk_jwstnircam_2024,
	title = {{JWST}/{NIRCam} {Transmission} {Spectroscopy} of the {Nearby} {Sub}-{Earth} {GJ} 341b},
	volume = {167},
	issn = {1538-3881},
	url = {https://doi.org/10.3847/1538-3881/ad19df},
	doi = {10.3847/1538-3881/ad19df},
	language = {en},
	number = {3},
	urldate = {2026-04-17},
	journal = {The Astronomical Journal},
	publisher = {The American Astronomical Society},
	author = {Kirk, James and Stevenson, Kevin B. and Fu, Guangwei and Lustig-Yaeger, Jacob and Moran, Sarah E. and Peacock, Sarah and Alam, Munazza K. and Batalha, Natasha E. and Bennett, Katherine A. and Gonzalez-Quiles, Junellie and López-Morales, Mercedes and Lothringer, Joshua D. and MacDonald, Ryan J. and May, E. M. and Mayorga, L. C. and Rustamkulov, Zafar and Sing, David K. and Sotzen, Kristin S. and Valenti, Jeff A. and Wakeford, Hannah R.},
	month = feb,
	year = {2024},
	pages = {90},
}

@article{alderson_jwst_2024,
	title = {{JWST} {COMPASS}: {NIRSpec}/{G395H} {Transmission} {Observations} of the {Super}-{Earth} {TOI}-836b},
	volume = {167},
	issn = {1538-3881},
	shorttitle = {{JWST} {COMPASS}},
	url = {https://doi.org/10.3847/1538-3881/ad32c9},
	doi = {10.3847/1538-3881/ad32c9},
	language = {en},
	number = {5},
	urldate = {2026-04-17},
	journal = {The Astronomical Journal},
	publisher = {The American Astronomical Society},
	author = {Alderson, Lili and Batalha, Natasha E. and Wakeford, Hannah R. and Wallack, Nicole L. and Aguichine, Artyom and Teske, Johanna and Adams Redai, Jea and Alam, Munazza K. and Batalha, Natalie M. and Gao, Peter and Kirk, James and López-Morales, Mercedes and Moran, Sarah E. and Scarsdale, Nicholas and Wogan, Nicholas F. and Wolfgang, Angie},
	month = apr,
	year = {2024},
	pages = {216},
}

@article{chen_asymmetry_2025,
	title = {Asymmetry and {Dynamical} {Constraints} in {Two}-limbs {Retrieval} of {WASP}-39 b {Inferring} from {JWST} {Data}},
	volume = {169},
	issn = {1538-3881},
	url = {https://doi.org/10.3847/1538-3881/adc803},
	doi = {10.3847/1538-3881/adc803},
	language = {en},
	number = {6},
	urldate = {2026-04-17},
	journal = {The Astronomical Journal},
	publisher = {The American Astronomical Society},
	author = {Chen, Zixin and Ji, Jianghui and Chen, Guo and Yan, Fei and Tan, Xianyu},
	month = may,
	year = {2025},
	pages = {294},
}

@article{moores_uv_2012,
	title = {{UV} degradation of accreted organics on {Mars}: {IDP} longevity, surface reservoir of organics, and relevance to the detection of methane in the atmosphere},
	volume = {117},
	issn = {2156-2202},
	shorttitle = {{UV} degradation of accreted organics on {Mars}},
	url = {https://agupubs.onlinelibrary.wiley.com/doi/10.1029/2012JE004060},
	doi = {10.1029/2012JE004060},
	language = {en},
	number = {E8},
	urldate = {2026-04-15},
	journal = {Journal of Geophysical Research: Planets},
	publisher = {John Wiley \& Sons, Ltd},
	author = {Moores, John E. and Schuerger, Andrew C.},
	month = aug,
	year = {2012},
}

@article{bardeen_improved_2013,
	title = {Improved cirrus simulations in a general circulation model using {CARMA} sectional microphysics},
	volume = {118},
	copyright = {©2013. American Geophysical Union. All Rights Reserved.},
	issn = {2169-8996},
	url = {https://onlinelibrary.wiley.com/doi/abs/10.1002/2013JD020193},
	doi = {10.1002/2013JD020193},
	language = {en},
	number = {20},
	urldate = {2026-04-15},
	journal = {Journal of Geophysical Research: Atmospheres},
	author = {Bardeen, C. G. and Gettelman, A. and Jensen, E. J. and Heymsfield, A. and Conley, A. J. and Delanoë, J. and Deng, M. and Toon, O. B.},
	year = {2013},
	note = {\_eprint: https://agupubs.onlinelibrary.wiley.com/doi/pdf/10.1002/2013JD020193},
	pages = {11,679--11,697},
}

@article{hartwick_high-altitude_2019,
	title = {High-altitude water ice cloud formation on {Mars} controlled by interplanetary dust particles},
	volume = {12},
	copyright = {2019 The Author(s), under exclusive licence to Springer Nature Limited},
	issn = {1752-0908},
	url = {https://www.nature.com/articles/s41561-019-0379-6},
	doi = {10.1038/s41561-019-0379-6},
	language = {en},
	number = {7},
	urldate = {2026-04-15},
	journal = {Nature Geoscience},
	publisher = {Nature Publishing Group},
	author = {Hartwick, V. L. and Toon, O. B. and Heavens, N. G.},
	month = jul,
	year = {2019},
	pages = {516--521},
}

@article{baumgarten_particle_2008,
	title = {Particle properties and water content of noctilucent clouds and their interannual variation},
	volume = {113},
	copyright = {Copyright 2008 by the American Geophysical Union.},
	issn = {2156-2202},
	url = {https://onlinelibrary.wiley.com/doi/abs/10.1029/2007JD008884},
	doi = {10.1029/2007JD008884},
	language = {en},
	number = {D6},
	urldate = {2026-04-14},
	journal = {Journal of Geophysical Research: Atmospheres},
	author = {Baumgarten, G. and Fiedler, J. and Lübken, F.-J. and von Cossart, G.},
	year = {2008},
	note = {\_eprint: https://agupubs.onlinelibrary.wiley.com/doi/pdf/10.1029/2007JD008884},
}

@article{le_feuvre_nonuniform_2008,
	title = {Nonuniform cratering of the terrestrial planets},
	volume = {197},
	issn = {0019-1035},
	url = {https://www.sciencedirect.com/science/article/pii/S0019103508001802},
	doi = {10.1016/j.icarus.2008.04.011},
	number = {1},
	urldate = {2026-04-09},
	journal = {Icarus},
	author = {Le Feuvre, Mathieu and Wieczorek, Mark A.},
	month = sep,
	year = {2008},
	pages = {291--306},
}

@article{minton_dynamical_2010,
	title = {Dynamical erosion of the asteroid belt and implications for large impacts in the inner {Solar} {System}},
	volume = {207},
	issn = {0019-1035},
	url = {https://www.sciencedirect.com/science/article/pii/S0019103509004953},
	doi = {10.1016/j.icarus.2009.12.008},
	number = {2},
	urldate = {2026-04-09},
	journal = {Icarus},
	author = {Minton, David A. and Malhotra, Renu},
	month = jun,
	year = {2010},
	pages = {744--757},
}

@article{thanathibodee_variable_2020,
	title = {Variable {Accretion} onto {Protoplanet} {Host} {Star} {PDS} 70},
	volume = {892},
	issn = {0004-637X},
	url = {https://doi.org/10.3847/1538-4357/ab77c1},
	doi = {10.3847/1538-4357/ab77c1},
	language = {en},
	number = {2},
	urldate = {2026-04-09},
	journal = {The Astrophysical Journal},
	publisher = {The American Astronomical Society},
	author = {Thanathibodee, Thanawuth and Molina, Brandon and Calvet, Nuria and Serna, Javier and Bae, Jaehan and Reynolds, Mark and Hernández, Jesús and Muzerolle, James and Hernández, Ramiro Franco},
	month = mar,
	year = {2020},
	pages = {81},
}

@article{machida_gas_2010,
	title = {Gas accretion onto a protoplanet and formation of a gas giant planet},
	volume = {405},
	issn = {0035-8711},
	url = {https://doi.org/10.1111/j.1365-2966.2010.16527.x},
	doi = {10.1111/j.1365-2966.2010.16527.x},
	number = {2},
	urldate = {2026-04-08},
	journal = {Monthly Notices of the Royal Astronomical Society},
	author = {Machida, Masahiro N. and Kokubo, Eiichiro and Inutsuka, Shu-ichiro and Matsumoto, Tomoaki},
	month = jun,
	year = {2010},
	pages = {1227--1243},
}

@article{johansen_forming_2017,
	title = {Forming {Planets} via {Pebble} {Accretion}},
	volume = {45},
	issn = {0084-6597, 1545-4495},
	url = {https://www.annualreviews.org/content/journals/10.1146/annurev-earth-063016-020226},
	doi = {10.1146/annurev-earth-063016-020226},
	language = {en},
	number = {Volume 45, 2017},
	urldate = {2026-04-08},
	journal = {Annual Review of Earth and Planetary Sciences},
	publisher = {Annual Reviews},
	author = {Johansen, Anders and Lambrechts, Michiel},
	month = aug,
	year = {2017},
	pages = {359--387},
}

@article{zawadzki_alma_2026,
	title = {The {ALMA} survey to {Resolve} {exoKuiper} belt {Substructures} ({ARKS}) - {III}. {The} vertical structure of debris disks},
	volume = {705},
	copyright = {© The Authors 2026},
	issn = {0004-6361, 1432-0746},
	url = {https://www.aanda.org/articles/aa/abs/2026/01/aa56505-25/aa56505-25.html},
	doi = {10.1051/0004-6361/202556505},
	language = {en},
	urldate = {2026-04-08},
	journal = {Astronomy \& Astrophysics},
	publisher = {EDP Sciences},
	author = {Zawadzki, B. and Fehr, A. and Hughes, A. M. and Mansell, E. and Kittling, J. and Han, Y. and Hou, C. and Pan, M. and Milli, J. and Olofsson, J. and Pearce, T. and Sefilian, A. A. and Nurmohamed, A. and Lee, J. and Mpofu, Y. and Bonduelle, M. and Booth, M. and Brennan, A. and Burgo, C. del and Carpenter, J. M. and Cataldi, G. and Chiang, E. and Ertel, S. and Henning, Th and Jankovic, M. R. and Kennedy, G. M. and Kóspál, Á and Krivov, A. V. and Lovell, J. B. and Luppe, P. and MacGregor, M. A. and Manamon, S. Mac and Marino, S. and Marshall, J. P. and Matrà, L. and Moór, A. and Pérez, S. and Weber, P. and Wilner, D. J. and Wyatt, M. C.},
	month = jan,
	year = {2026},
	pages = {A197},
}

@article{kenyon_prospects_2005,
	title = {Prospects for {Detection} of {Catastrophic} {Collisions} in {Debris} {Disks}},
	volume = {130},
	issn = {1538-3881},
	url = {https://iopscience.iop.org/article/10.1086/430461},
	doi = {10.1086/430461},
	language = {en},
	number = {1},
	urldate = {2026-04-08},
	journal = {The Astronomical Journal},
	publisher = {IOP Publishing},
	author = {Kenyon, Scott J. and Bromley, Benjamin C.},
	month = jul,
	year = {2005},
	pages = {269},
}
\bibliographystyle{aasjournalv7}



\end{document}